# Influence of Surface Roughness on Linear Behavior and Mechanical Properties of Three Cyanoacrylate-Based Adhesives Used to Bond Strain Gages

L. G. Simão[a,*], W. P. Jesus[a], M. E. A. Ribeiro[a], H. C. Rangel[a], R. J. S. Rodríguez[a], E. A. Carvalho[a]

[a]Northern Fluminense State University, Advanced Materials Laboratory, Brazil

## Abstract

The challenge of accessing specialized adhesives designed for strain gage applications has been highlighted due to failures in logistic chains, requiring the exploration of local alternatives. A direct simulation of strain gage bonding behavior with two steel plates is infeasible due to the unique construction of strain gages. Therefore, an indirect simulation method, comparing local alternatives to a widely accepted adhesive, Loctite 496®, was employed in this study. Two potential replacements, Loctite 401® and Tekbond 793®, were tested and matched against the benchmark adhesive, with a focus on the key mechanical properties: Proportional Shear Strain (PSS), Proportional Shear Stress (PSSt), and Apparent Shear Modulus (G*). Loctite 401® exhibited the highest G*, suggesting its potential use in strain gage installations if G* is considered most important. However, Tekbond 793® demonstrated superior PSS, Maximum Shear Stress (MSSt), and Rupture Shear Stress (RSSt) performance, displaying linear behavior even without an accelerator. Surface preparation considerations were also discussed, noting that hand abrading results in double the surface roughness than using an orbital sander. The study further identified two main regions concerning failure modes related to Ra, with values below 0.31 μm causing significant variations in observed mechanical properties, pointing towards factors beyond adhesive layer thickness affecting bond properties. Lastly, the general recommendation is the use of an accelerator for all tested adhesives, while the use of a surface conditioner and neutralizer was found to negatively impact adhesive performance.

**Keywords:** Lapshear test, Strain Gage bonding, Roughness, Industrial Adhesive, Mechanical Properties

## 1. Introduction

Bonded strain sensors are extensively used in diverse settings, from sterile laboratories to extreme field applications. This versatility comes with a range of high-performance adhesives, designed specifically to transmit strains to the sensor as accurately and sustainably as physically possible. However, to achieve this, the surface must have a level of roughness that enables the sensor to perform optimally, rather than the adhesive. Various abrasion methods, used either independently or in combination, can generate the desired surface finish. For instance, Vishay's literature [1] recommends the following for strain gages (Table 1):

**Table 1**
Prescribed surface finish for Strain Gage installation [1]

| Class of Instalation | Surface Finish, rms | |
|---|---|---|
| | Min | µm |
| General Stress Analysis | 63 – 125 | 1.6 – 3.2 |
| High Elongation | >250 | >6.4 |
| | Cross-Hatched | |
| Transducers | 16 - 63 | 0.4 – 1.6 |

The aforementioned surface finishes fall below the 3–5 µm range typically recommended by adhesive manufacturers for grit blasting [2]. In some cases, a sensor may be bonded to a surface following only chemical cleaning and neutralization, as in the case of chromed axles, electronic circuits, and similar components. It is also important to comprehend the limitations of cyanoacrylate adhesives in situations where the surface roughness is below the levels outlined in Table 1.

### *1.1 Strain Gage Bonding Adhesives*

The adhesive used to bond a strain gage to a part, machine, or material plays a vital role in the measurement process, as a poorly performing adhesive can compromise even the most optimal gage installation and signal conditioning equipment. It is also well known that no single adhesive is ideally suited for all applications, thus several glues are listed as possible solution for strain gage installation. Sensor manufacturers usually provide their own bonding system [3]. However, these adhesives often have a shelf life of approximately a year, require high procurement costs, and are not easily available worldwide. Consequently, many users seek local and suitable substitutes. The most common adhesives are cyanoacrylate-based and epoxy-based. The former is used for routine stress analysis under mild environmental conditions, curing in a few minutes to form what is considered a creep-free, fatigue-resistant bond with high elongation capability [1]. However, the performance of this adhesive can be easily degraded over time by high humidity, moisture absorption, and elevated temperature. This requires the coating of the strain gage installation with a protective layer for long-term applications. Epoxy-based adhesives while more versatile and suitable for harsher environments, may require curing at elevated temperatures and careful component mixing [3]. This study benchmarks an available strain gage adhesive replacement, Loctite 496®, recommended by a local sensor seller, along with two other comparable ones that share similarities: Loctite 401® and Tekbond 793®.

### *1.2 Cyanoacrylate-Based Adhesives*

The first two commercial cyanoacrylate-based adhesives are widely employed to bond strain sensors to diverse surfaces under different environmental conditions, ranging from laboratory to field applications. The third adhesive in Table 2 is a multipurpose glue, occasionally used as a temporary solution for sensor installation when high-performance adhesive is unavailable. The data presented in Table 2, collected from available manufacturers' literature, summarize the relevant information.

**Table 2:**
Studied adhesives. Manufacturers' data.

| Adhesive | Density ($g/cm^3$) | Viscosity (mPa.s) | Lap Shear Strength (MPa)* | Key Substrates | Chemical Base |
|---|---|---|---|---|---|
| 496® [4] | 1.10 | 70 to 120 | 20 to 30 | Metals, Plastics and Elastomers | Methyl Cyanoacrylate |
| 401® [5] | 1.06 | 70 to 110 | 17 to 24 | Metals, Plastics and Elastomers | Ethyl Cyanoacrylate |
| 793® [6] | 1,04 | 80 to 120 | > 10 | Porous Materials, Plastics and elastomers | Ethyl Cyanoacrylate |

* - steel substrate

### *1.3 Stress Distribution and Failure Characteristics*

The stress distribution in single lap shear joints is well-understood and has been extensively studied [7,8,9]. According to [9], a flexible adhesive applied in a very thin layer is subjected to higher stresses at the joint edges, where failure is expected to initiate. Bending, however, is entirely disregarded due to the adhesive layer being several orders of magnitude thinner than the steel strips used. For adhesive layers thinner than 1 mm, the following expression is regarded as a conservative estimate:

$$P_{gy} = \tau_y b t \tag{1}$$

Where $P_{gy}$ is the maximum load the section can bear, $\tau_y$ refers to the shear yield strength of the adhesive, *b* represents the joint width, *t* denotes the overlap length. Figure 1 presents a schematic representation of the shear stress versus shear strain curve.

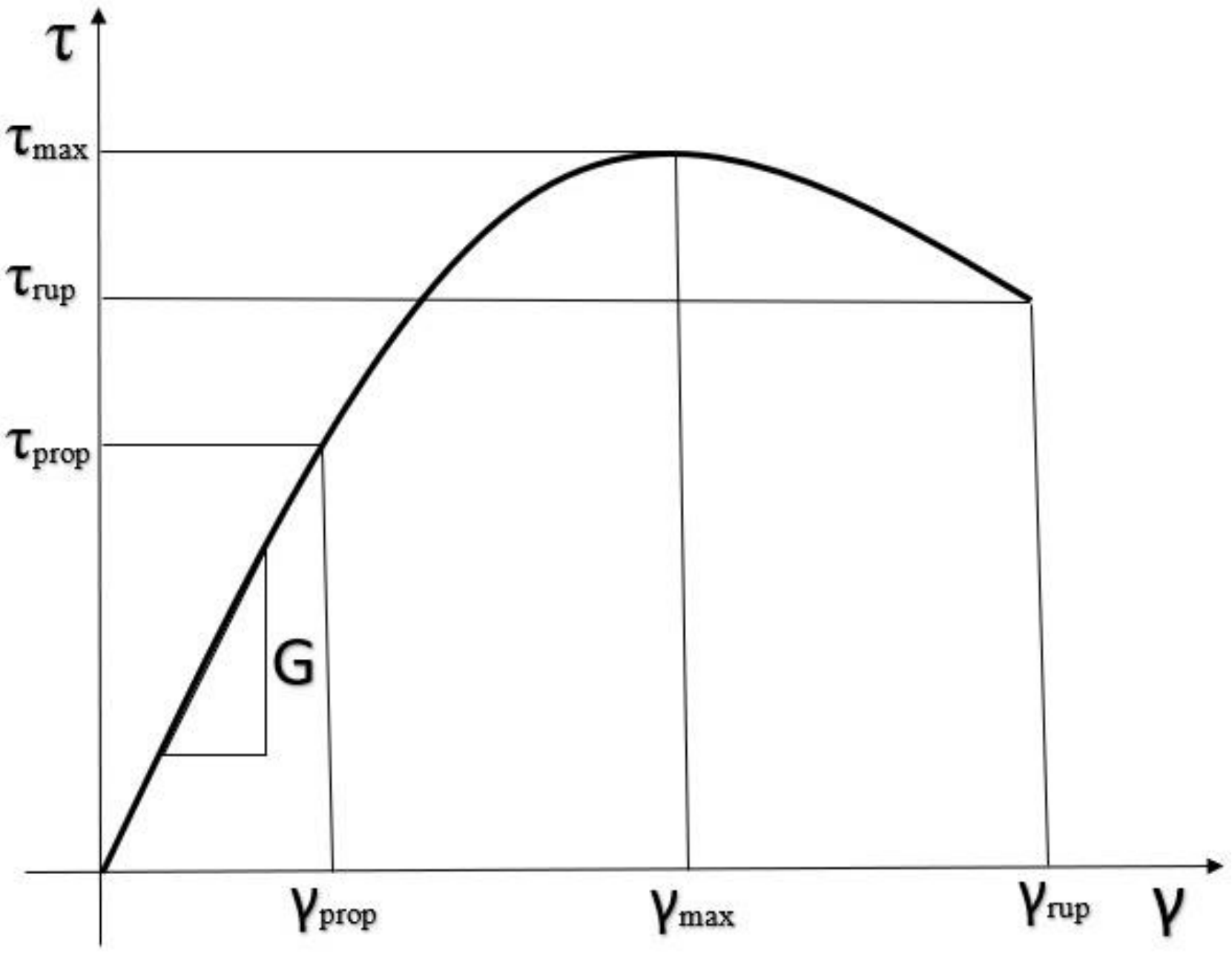

**Fig. 1.** Shear stress versus shear strain curve schematics.

Several properties can be determined from this curve, such as: Maximum Shear Stress, $\tau_{máx}$, is the maximum stress value reached by the lap shear test and also indicates the point beyond which the adhered joint stops functioning. While it is an important feature for general adhesive usage, it falls outside the linear area for sensor installation and is thus not recommended. Rupture Shear Stress, $\tau_{rup}$, corresponds to the stress value at which a total separation of the metal strips is observed during the lap shear test. Beyond this point, the joint is no longer useful. Proportional Limit Stress, $\tau_{prop}$, represents the maximum linear stress value. This is the ideal working stress range, since the load can be applied and removed without causing significant damage to the joint. Shear Modulus

(G) provides a linear correlation between shear stress and shear strain. However, as reported in [10], the true value of G cannot be determined by a lap shear test alone. Instead, they present four scenarios where G is established using Finite Element Analysis (FEA) and subsequently contrasted with experimental results. The results revealed that the FEA-determined G remains constant, while the experimental G decreases with increasing adhesive thickness. For adhesive layers varying between 0.5 and 3.0 mm, G varies almost linearly. For thicknesses less than 0.5 mm, the value of G increases stepwise. Moreover, the larger the simulated G, the steeper the variation for thinner layers because below a certain thickness, the stress distribution caused by non-uniformity in the overlapping region becomes dominant in the stress distribution. Maximum shear strain ($\gamma_{máx}$) is the maximum strain value reached by the lap joint before it collapses. Rupture Shear Strain ($\gamma_{rup}$) is the strain value at which total separation of the metal strips occurs. Proportional Limit Strain ($\gamma_{prop}$) is the maximum value at which the strain remains linear. The Shear Spring Constant (k) represents the force-to-distance ratio of the lap joint. This auxiliary property is used to verify that the clip gage measures deformation primarily due to the significant adhesive deformation, not the strain associated with the adherent. The total displacement can be calculated by multiplying the recorded strain value in the overlap region by the clip gage length (25 mm). A comparison of k values for the overlap region and steel strips ($k_{steel}$ = 10000 kN/mm) reveals that a k value several orders of magnitude lower than $k_{steel}$ indicates that the adhesive displacement is the only measurable portion.

## 2. Materials and Methods

### *2.1 Surface Preparation and Roughness Measurement*

For this study, six pairs of SAE 5160 steel strips, each measuring 150 x 25 x 3, were cut and abraded using seven distinct sandpaper grain sizes, with only one side of each strip subjected to testing. As most strain gage surface preparation is performed manually, this study introduced a manual test where two individuals – Analyst A, an experienced strain analyst, and Analyst B, an undergraduate engineering student – each abraded two strips of identical steel. The same sandpaper brand was used, with grit numbers #120, #220, and #320.

The materials were employed in their “as-received" condition, and a series of ten hardness tests indicated a value of 24 $H_{Rc}$. The strips were collectively positioned on a flat surface, encased by 'sacrificial' plates, and secured using double-sided tape. Subsequently, sandpaper was affixed to a circular orbital sander, employed to create a uniform surface in terms of roughness. The use of this particular sander can be attributed to its inherent capacity to generate consistent roughness in all directions due to its random oscillatory motion, a feature not found in manual abrasion, especially when performed in confined spaces. This characteristic enhances the reproducibility and control of the testing process. Once the surface achieved a satisfactory visual standard, the strips were removed, cleaned with neutral soap and running water, and immediately dried using a heat gun. The samples that have undergone solely this process will henceforth be denoted as "Simply Abraded" or SA. This method represents the most common approach for sensor bonding in field applications. Following this preparation, the strips were arranged on a flat surface and the roughness was measured using a Mitutoyo SJ-201P Surface Roughness Tester. Although this test is influenced by the ASTM D5656-17, it does not strictly conform to the standard [11]. As six strips were used in each particular test and 12 per roughness total, 30 and 60 total measurements were made each time.

It should be noted that in the laboratory and transducer industries, it is a common practice to perform some form of chemical treatment on the surface prior to sensor bonding. Therefore, the present study examines two such treatments, both independently and in conjunction, and reports on their effects on one of the adhesives.

The first treatment involves a combination of oxidation, oils, grease, and contaminant removal, typically referred to as a "conditioner". This is a mild phosphoric acid solution, which acts as a gentle etchant and expedites the cleaning process [1]. The solution is applied using cotton swabs in unidirectional motions, and any excess is subsequently removed with a clean paper towel. It is important to note that this solution cannot be used independently due to its potential to alter the surface pH and subsequently impact the bonding ability of cyanoacrylates. The application of the conditioner necessitates the subsequent use of a solution to return the pH to a neutral state, referred to as a "neutralizer". This is an ammonia-based solution, capable of neutralizing any chemical reactions initiated by the conditioner, with the objective to create optimum surface conditions for most strain gage adhesives [1]. The application mirrors that of the conditioner, utilizing unidirectional swab strokes and a thorough drying process at the end. In this study, this surface treatment, when applied, will be referred to as "Conditioner + Neutralizer" or simply “CN”.

The additional surface treatment employed involves the use of a catalyst or "accelerator". Numerous manufacturers supply such catalysts designed to hasten the bonding process of cyanoacrylate adhesives. For this experiment, Turbo Primer®, a product by Garin Brasil, and available in aerosol cans, was selected as the catalyst. After its application, it was allowed to dry outside the laboratory to ensure the dissipation of any residual fumes. Once dried, it was ready for use.

The bonding process for all specimens followed the same methodology: after appropriate surface preparation, an adhesive layer was applied to the abraded region on both strip surfaces. The adhesive-coated faces were subsequently aligned, and thumb pressure was applied to facilitate bonding. During this process, the strips were laterally aligned by using a flat surface as a reference. Once initial bonding was achieved, the specimen was positioned on its 125 x 25 side on the same flat surface. A "C" clamp, equipped with an eraser serving as a pressure pad, was then used to apply sustained pressure. This setup was maintained for 24-hour to allow for the complete curing of the adhesive. Subsequently, the specimen was considered prepared and suitable for tension testing.

### *2.2 Lap Shear Specimen Preparation and Test*

The lap shear test is a well-established methodology, covered by standards such as ASTM D5656 [11]. Although our interest deviates from assessing the adhesive performance associated with a grit-blasted surface as the substrate, the focus of our study leans towards finely sanded surfaces exhibiting significantly flatter roughness compared to those typically required by usual tests. The specimen consisted in two AISI 5160 cutlery steel strips, each having dimensions of 100 x 25 x 3 mm, where one end of each strip was prepared as described. Additional short blanks were cut from the same material to ensure load alignment, a pure shear stress state, and the appropriate placement for clip gage loading. For shear strain measurement, an Instron sensor was used, with data recording occurring simultaneously with the load. The bonded area was designed to fit within the clip gage displacement range (25 ± 3 mm). The sensor was carefully placed, secured with steel clamps and rubber bands, thereby ensuring that no slippage occurred during the testing process (Figure 2).

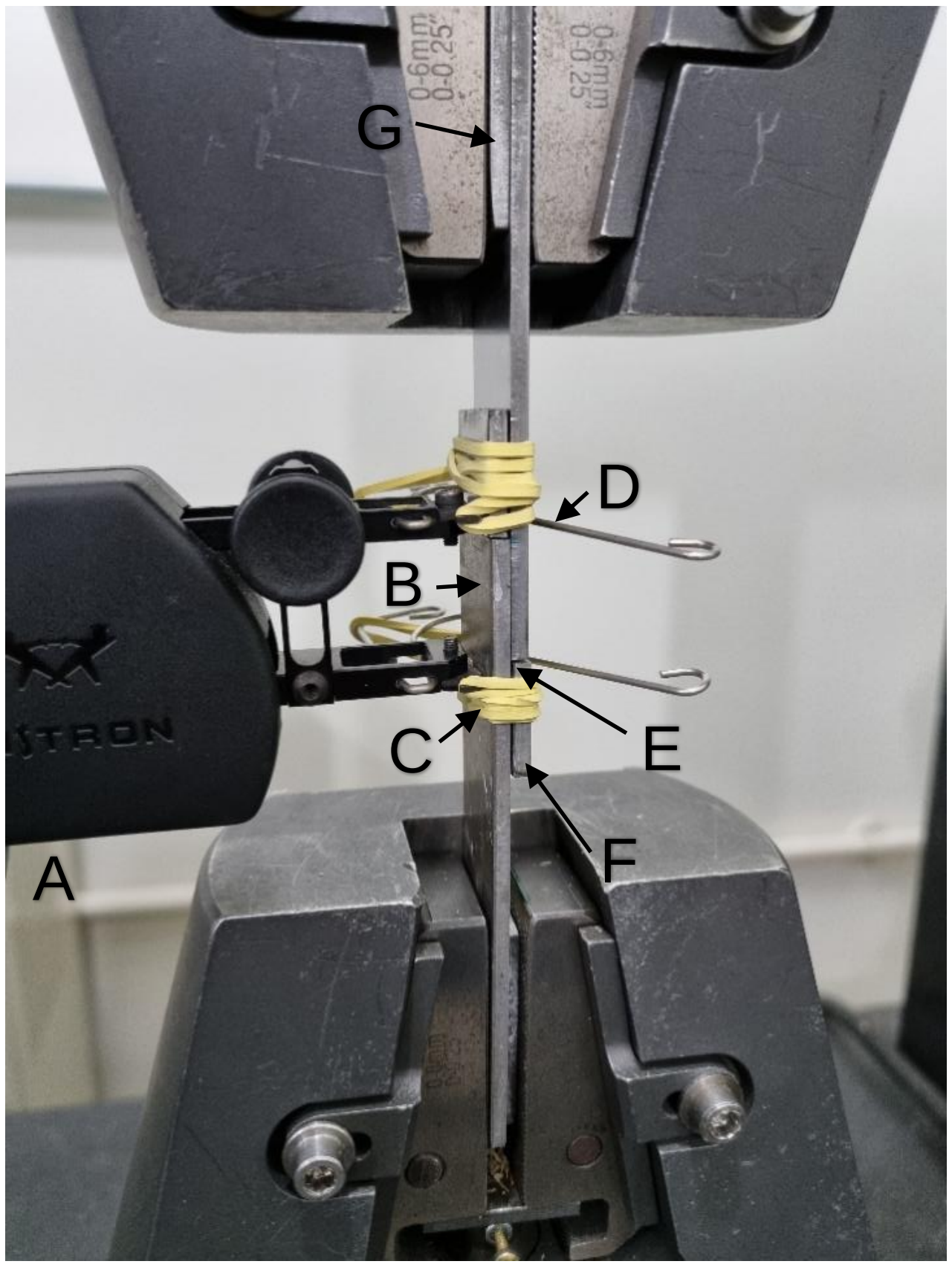

**Fig. 2.** Schematic representation of a lap shear test specimen. (A) Instron clip gage, (B) Lap shear joint, (C) Rubber band for securing the sensor, (D) Clip gage positioning clamp, (E) Gap in the alignment, (F) Clip gage alignment plate, and (G) Test alignment insert.

Figure 3 presents a scheme explaining the lap shear joint shown in Figure 2. As previously discussed, due to the layer thickness, the bending component is not noticeable. The two smaller inner blanks are bonded together to ensure a parallel stance for the clip gage (Figure 2A), thus enabling clear readings.

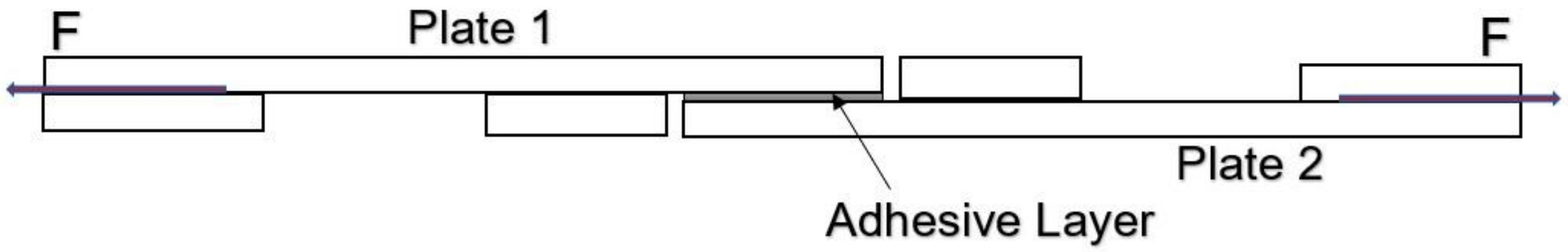

**Fig. 3**. Sample schematics.

*2.3 Deformation Reading and Conversion to Shear*

The clip gage in this study used was originally designed for tensile testing; therefore, post-collection data conversion is required. Figure 4 shows the samples and test schematics.

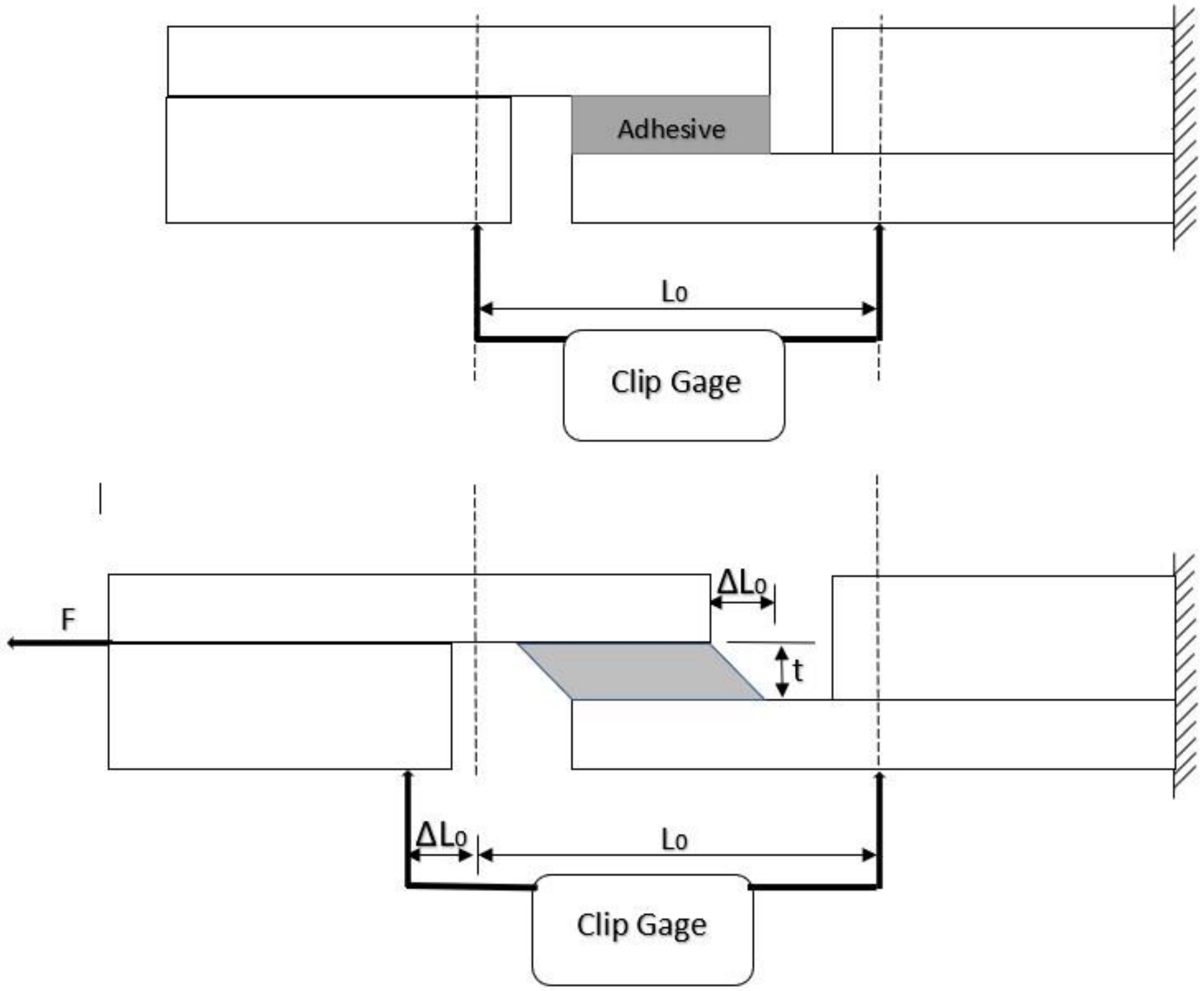

**Fig. 4.** Experimental setup schematics. Drawing is not made to scale.

The experimental setup employed has one side fixed, similarly to a typical testing machine. This arrangement ensures that the overall displacement measured is identical to the displacement applied to the adhesive layer, without affecting the thickness. One of the jaws fixing the sample is attached to the test machine frame, remaining stationary while the other moves along the loading line. The clip gage is attached and calibrated to an initial gage length of 25.23 mm (between contact knives). The recorded strain is not shear strain and should not be used as such. Expression (2) describes the aforementioned relationships and allows for the determination of the variation $\Delta L_0$ occurring during the test:

$$\varepsilon = \frac{\Delta L_0}{L_0} \therefore \Delta L_0 = \varepsilon L_0 \tag{2}$$

Once this variation in the initial length is determined, and the adhesive layer thickness is measured (a process that can only be performed post-testing when the metal plates have been separated), the shear strain ($\gamma$) can ultimately be calculated. This involves combining the experimental data obtained from the clip gage with the relationships defined in expressions (2) and (3).

$$\gamma = \frac{\Delta L_0}{t} \tag{3}$$

Where, $\gamma$ represents the shear strain and $t$ denotes the thickness of the adhesive layer.

### *2.4 Surface and Chemical Enhancement Effect Over Shear Stress and Strain*

Several sensor and strain gage manufacturers recommend surface neutralization and the chemical accelerator use to improve the durability and performance of the bonded interface, thus ensuring reliable data over time. As previously described, the surface conditioning process eliminates potential oxidation, and subsequently, a neutralizing solution is applied. The bonding process is promptly conducted thereafter. The accelerator is recommended to be applied post-neutralization of the surface, such as on the rear face of the strain gage. The Loctite 496 adhesive, widely used as a substitute, was thus subjected to all four aforementioned surface preparation conditions.

## 3. RESULTS, VALIDITY AND DISCUSSION

### *3.1 Surface Roughness*

The results are detailed in Table 3. Figure 5 shows the results per sandpaper grit (#n), with the error bars representing a 95% confidence interval.

**Table 3**
Average surface roughness ($R_a$) and its sample standard deviation corresponding to each sandpaper grit (#n) used on the AISI 5160 steel surface.

| Sandpaper Grit | 36 | 50 | 80 | 120 | 220 | 320 | 400 |
|---|---|---|---|---|---|---|---|
| $R_a$ (µm) | 1.08 | 0.85 | 0.31 | 0.21 | 0,17 | 0,12 | 0.07 |
| Std Dev $R_a$ (µm) | 0,10 | 0,09 | 0,05 | 0,02 | 0,02 | 0,02 | 0,01 |

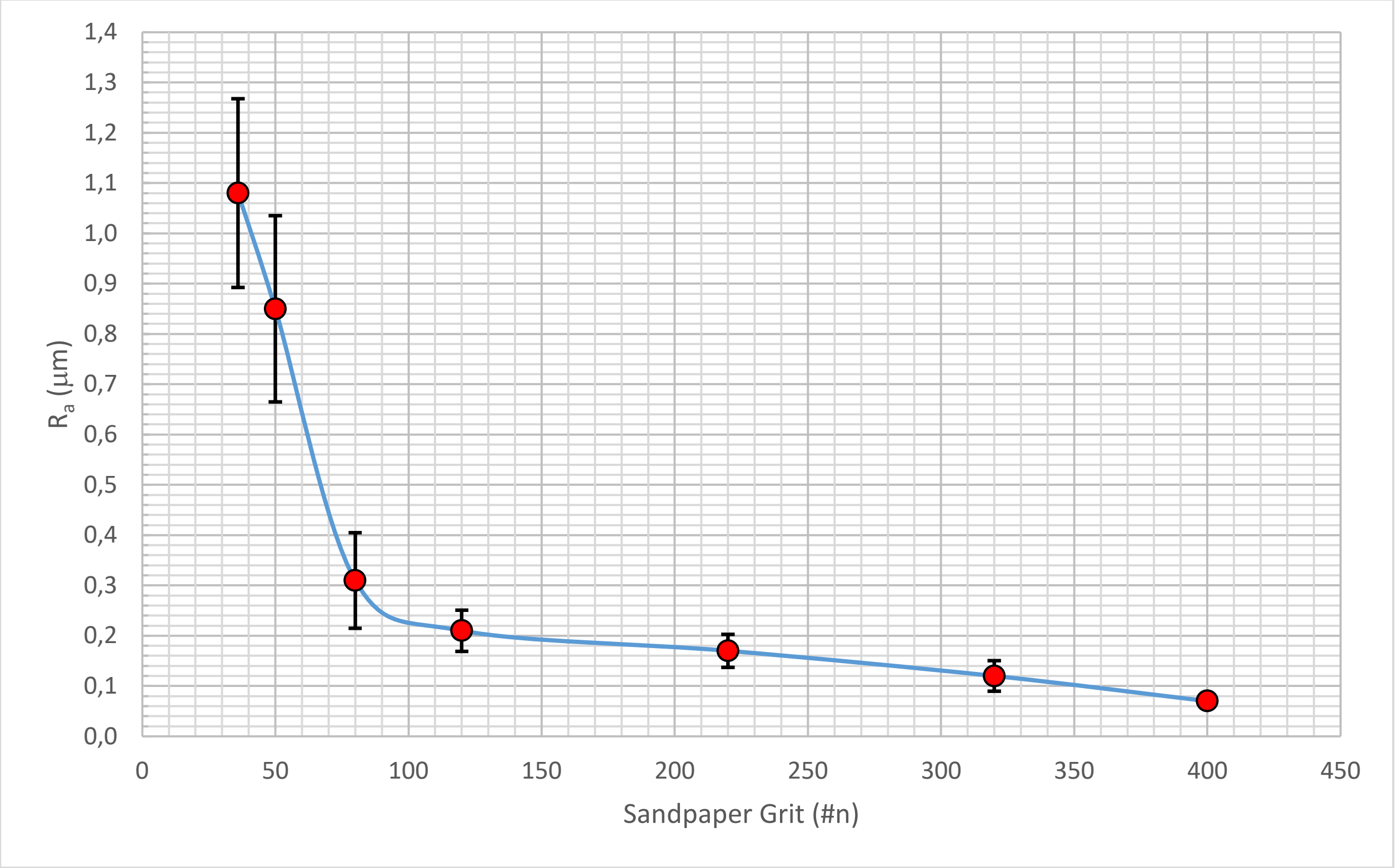

**Fig. 5.** Surface roughness ($R_a$) of each sandpaper grit (#n) on AISI 5160 steel surface, with 95% confidence intervals.

The results shown in Table 4 indicate that while experience plays a minimal impact on the final outcome, manual abrasion produces coarser surfaces than those conducted by orbital sanders.

**Table 4**
Roughness produced by hand abrading, compared to mechanical sanding.

| | #120 | | | | #220 | | | | #320 | | | |
|---|---|---|---|---|---|---|---|---|---|---|---|---|
| | Long | | Trans | | Long | | Trans | | Long | | Trans | |
| | Av | Std Dv | Av | Std Dv | Av | Std Dv | Av | Std Dv | Av | Std Dv | Av | Std Dv |
| **Analyst A** | 0.34 | 0.03 | 0.30 | 0.02 | 0.28 | 0.02 | 0.24 | 0.03 | 0.22 | 0.01 | 0.17 | 0.04 |
| **Analyst B** | 0.35 | 0.02 | 0.34 | 0.02 | 0.30 | 0.01 | 0.26 | 0.03 | 0.20 | 0.01 | 0.17 | 0.02 |
| **Analyst A (Avg)** | **0.32** | | | | **0.26** | | | | **0.20** | | | |
| **Analyst B (Avg)** | **0.34** | | | | **0.28** | | | | **0.19** | | | |
| **Orbital Sander** | **0.21** | | | | **0.17** | | | | **0.12** | | | |

Conventional practice typically designates #220 or #320 as the final sandpaper before bonding a strain gage to a steel surface. This study, however, also included #120 grit in the process, acknowledging the often-recommended practice of initially abrading with a coarser grit prior to the final one.

### *3.2 Modes of Failure*

Following each test, the failure mode was evaluated, revealing two major spatial distribution aspects and two thickness ratios. They were combined to produce the four failure modes observed in this study all of which were considered valid and are in alignment with existing literature [12]. Additionally, all adhesive types demonstrated these failure modes without any preferential pattern. These aspects are further demonstrated in Figures 6–8.

The darker areas indicate a thicker adhesive layer. Regarding the spatial distribution of thicker versus thinner adhesive layers, two possibilities were observed: an even distribution (Figure 6) and a dominant side (Figure 7).

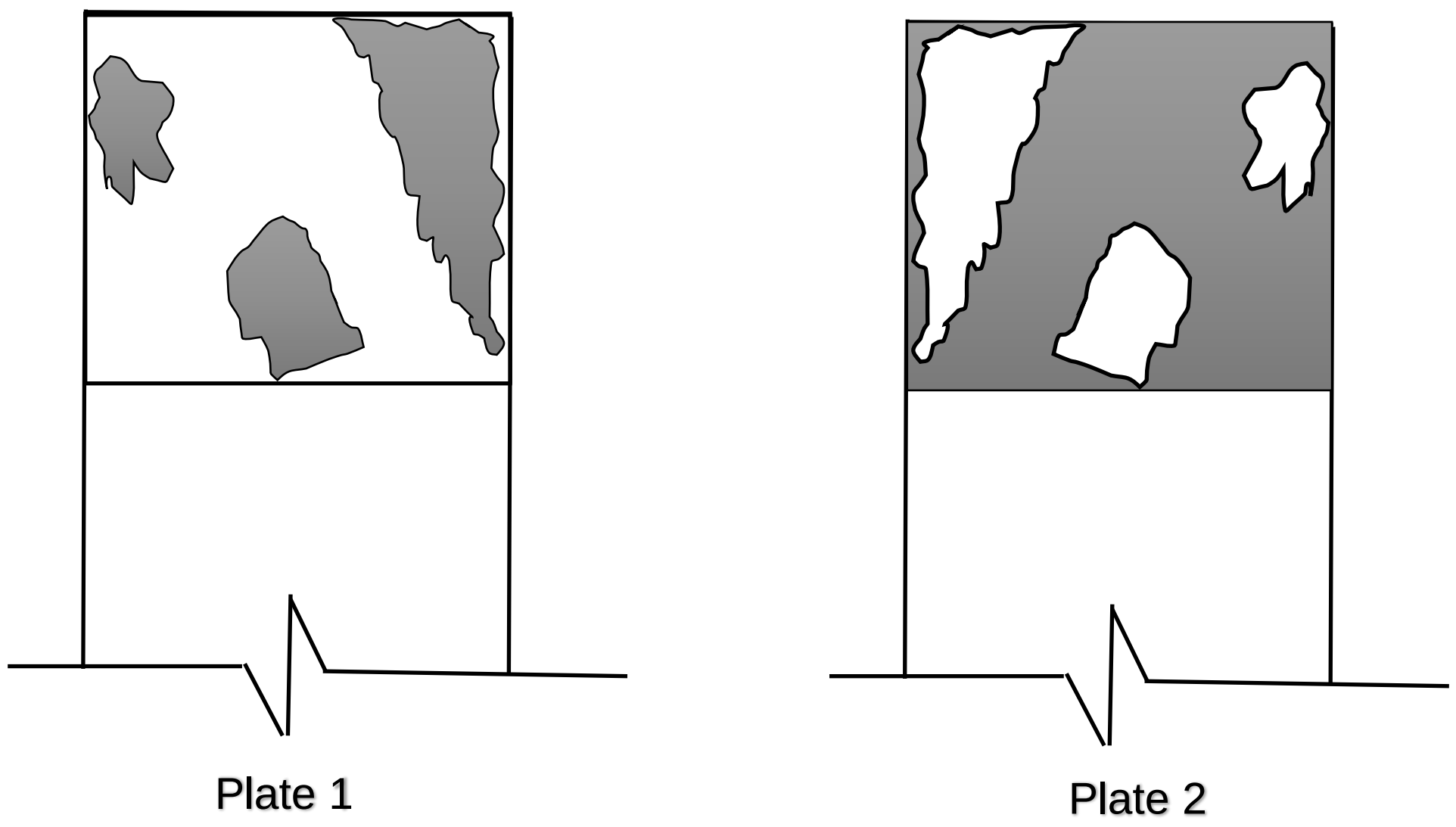

**Fig. 6.** Failure mode type I where there is no dominant side retaining the largest area of the glue (darkest part).

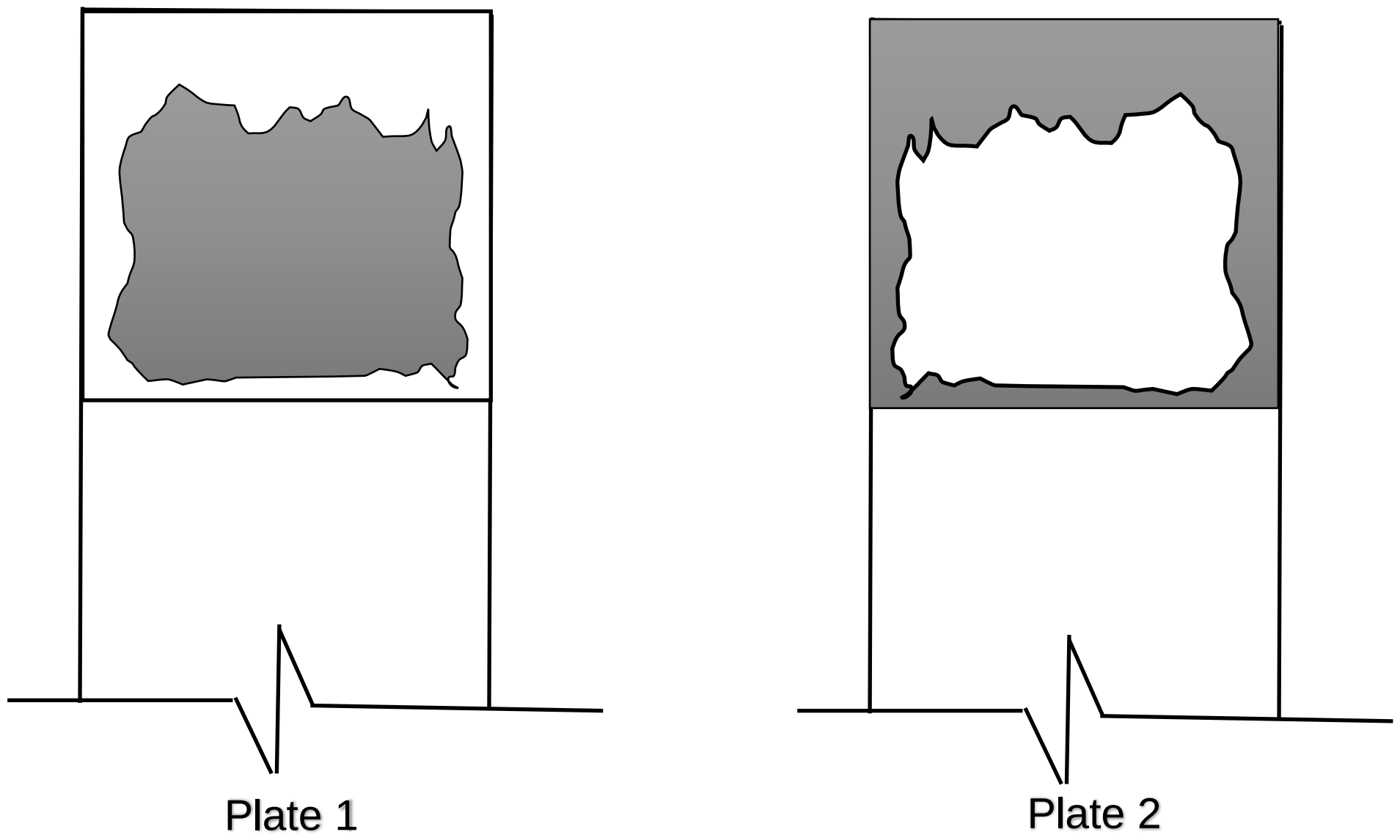

**Fig. 7.** Failure mode type II where there is a dominant side retaining the largest area of the glue (darkest part).

It should be noted that this is a subjective criterion, based on visual inspection, unlike thickness measurements which is objectively quantified. As to the thickness, a typical adhesive failure was around the middle of the bonding layer, while another common occurrence was the formation of two uneven sides (Figures 8A and 8B).

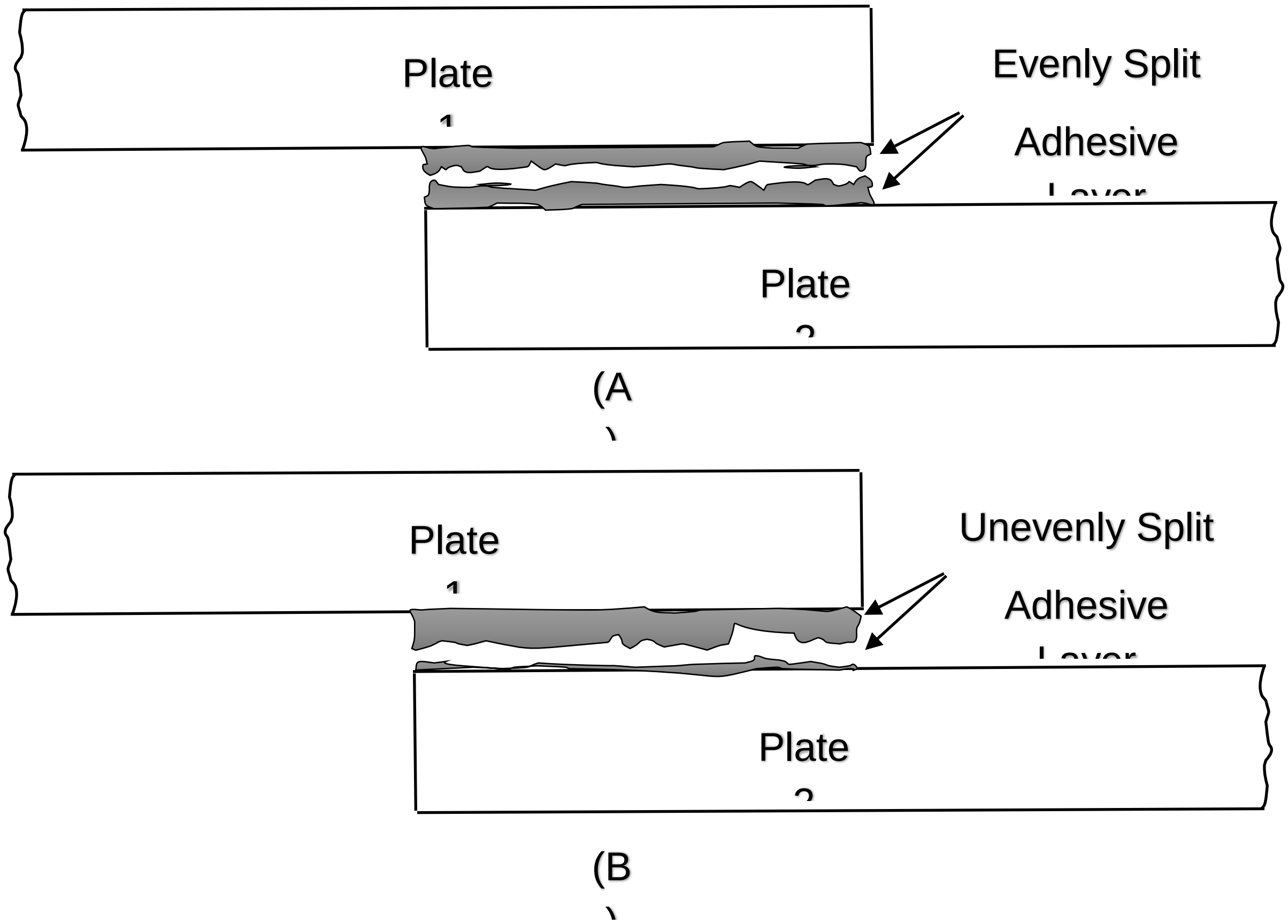

**Fig. 8.** Failures modes thickness types: the first splits the adhesive layer in half, and the second generates an uneven thickness between the two halves. (Not to scale).

Table 5 shows the average values for the collected data. Layer thickness was measured using a Mitutoyo Digital Layer Measurer Digi-Derm Series 979 (1μm resolution).

**Table 5**
Average adhesive layer thickness, with and without the use of an accelerator.

| Adhesive | Thickness (μm) |
| --- | --- |

| | Plate 1 (avg) | Std Dev 1 | Plate 2 (avg) | Std Dev 2 | Total (avg) |
|---|---|---|---|---|---|
| Loctite 496 | 14 | 5,6 | 15 | 5,9 | 29 |
| Loctite 496 + Turbo Primer | 19 | 7,2 | 15 | 5,1 | 34 |
| Loctite 401 | 23 | 10,8 | 16 | 9,1 | 39 |
| Loctite 401 + Turbo Primer | 20 | 13,6 | 21 | 11,9 | 41 |
| Tekbond 793 | 15 | 5,8 | 15 | 7,2 | 29 |
| Tekbond 793 + Turbo Primer | 19 | 7,0 | 19 | 7,5 | 38 |

According to MatWeb website [13], cyanoacrylate adhesive layer thickness averages around 45 μm, ranging between 20 and 100 μm. All values observed in this study fell within this range, although below the stated average.

### *3.3 Roughness and Grit Effect Over Apparent Shear Modulus (G*)*

The distinction between the True Shear Modulus (G) and the Apparent Shear Modulus (G*) arises from the adhesive layer thickness being insufficient for an accurate measurement of this property, especially as this is not measured directly over a shear specimen, such as in an *Iosipescu* test [14]. According to [7] the true G value cannot be determined by lap shear test. Instead, it presents four cases in which G is calculated using Finite Element (FE) analysis, then compared to experimental values. The FE approach treated G as a constant, whereas the experimental G decreased with an increase in thickness. Interestingly, the G variation exhibits linearity between 0.5 and 3.0 mm. However, for values below 0.5 mm, a significant increase was observed. Moreover, the larger the simulated G, the steeper was the variation across thinner layers. It should be noted that below a certain level of thickness, the stress distribution triggered by non-uniformity within the overlap region becomes a dominant factor in the overall stress distribution. Therefore, it is expected that the actual G may be smaller than the values recorded in this research.

In this study, G* is also analyzed in relation to variations in sandpaper grit number (#n), given the practicality of associating a specific property with a grit number as opposed to a $R_a$ value. It is essential to remember that all results are specifically valid when associated with a steel that exhibits hardness of 24 $H_{Rc}$.

Figure 9 provides the results for $R_a$ and G*, while Figure 10 displays the same data but categorized by sandpaper grit number. Loctite 401® is the stiffest adhesive, achieving a peak at 0.21 μm. Tekbond 793® yielded results comparable to the benchmark for practical grits (#120 and beyond). It is noteworthy that for Loctite 496, the optimal $R_a$ for maximum stiffness is found between 0.17 and 0.21 μm, corresponding to grits #220 and #320. A comprehensive set of values is provided in Appendix C.

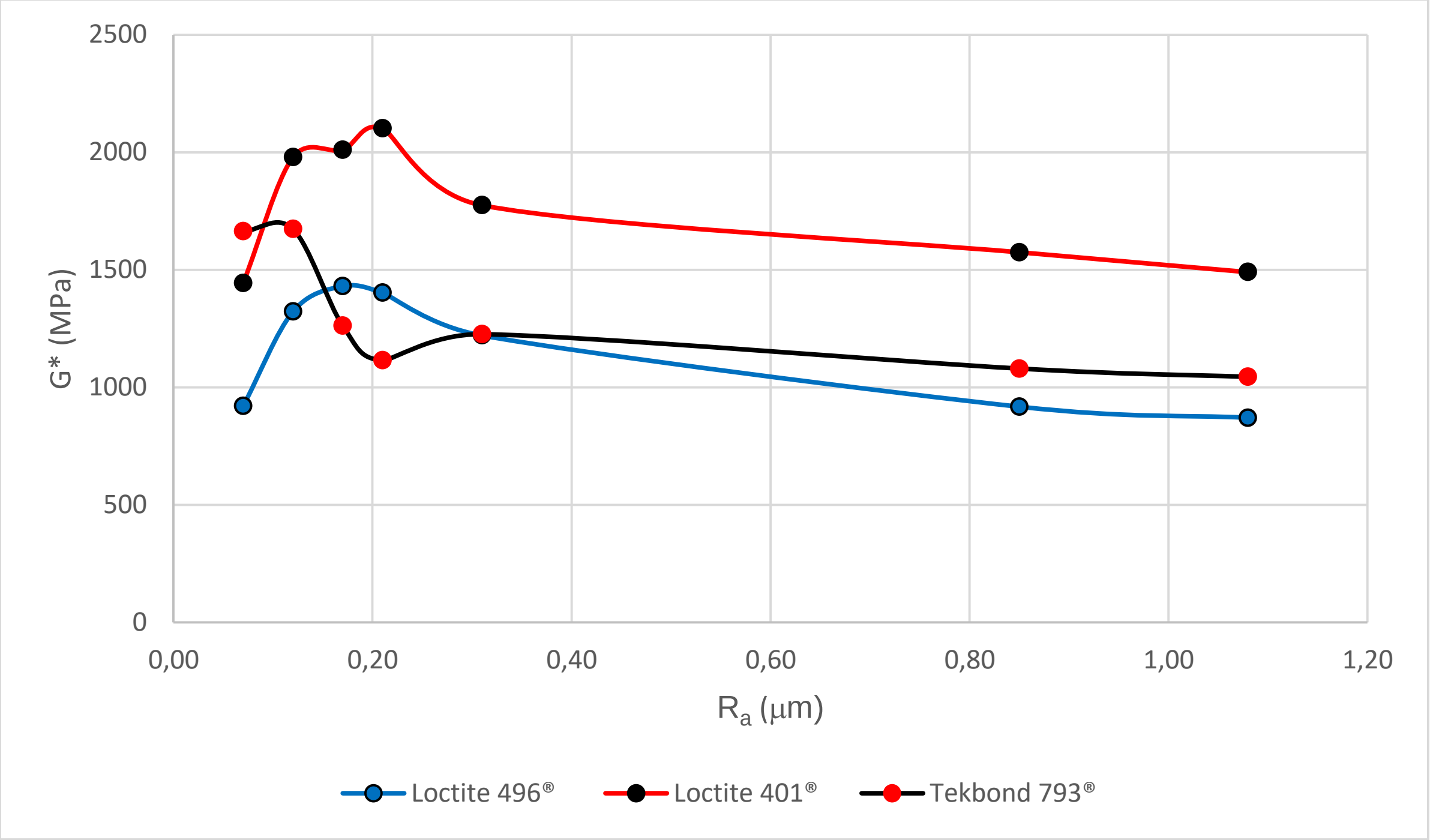

**Fig. 9.** Apparent Shear Modulus x $R_a$

In regard to Figure 9, the estimated value for G is given by [13] and calculated using the E range (1.36 – 3.07 GPa) and $\nu$ = 0.35, which results in G = 500 → 1140 MPa.

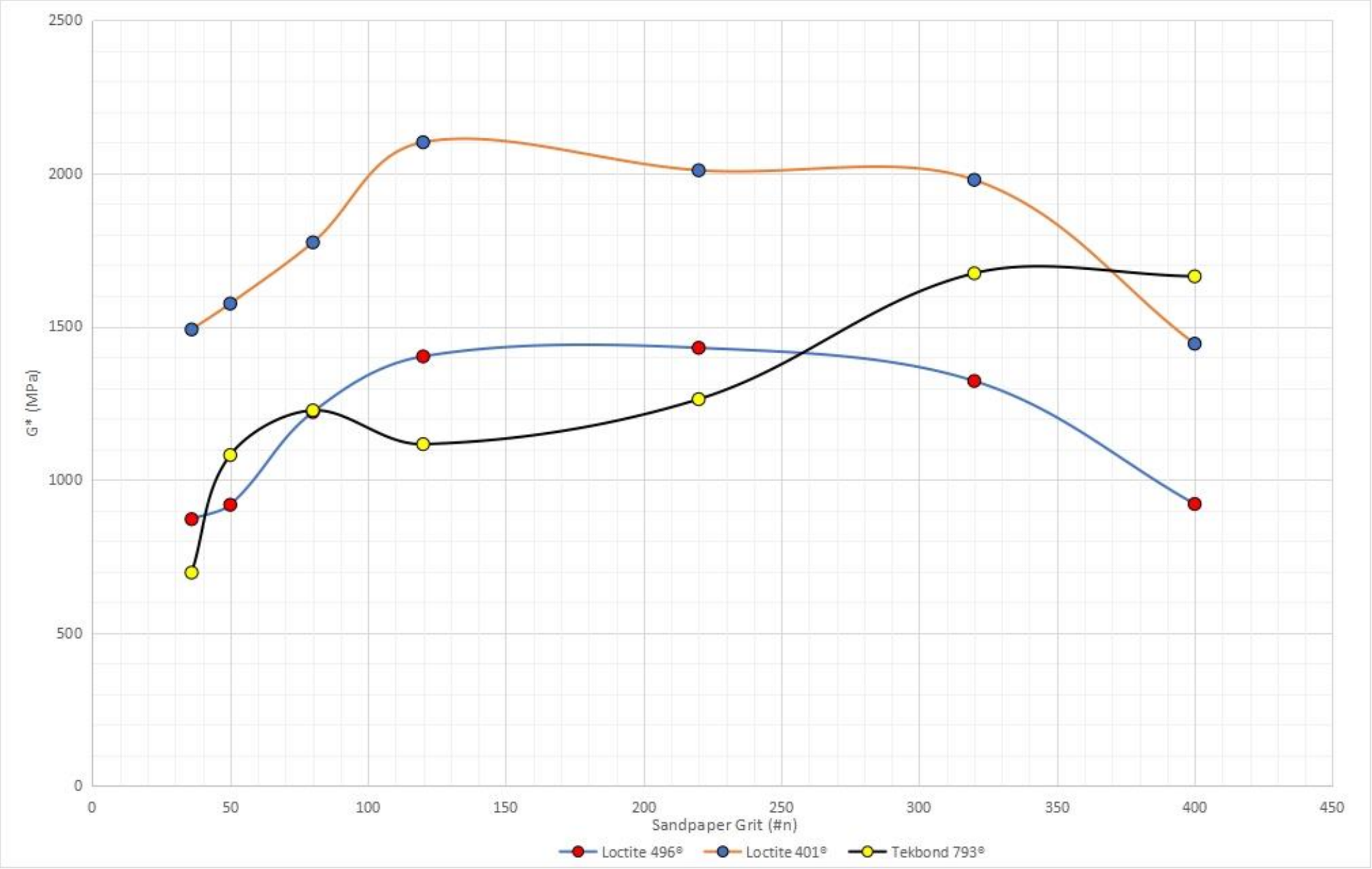

**Fig. 10.** Apparent Shear Modulus x Sandpaper Grit (#n)

*3.4 Proportional Shear Strain versus $R_a$ and Grit*

The Proportional Shear Strain (PSS) is perhaps the most sought-after mechanical property among strain gage users, as it indicates the operational limit for a given adhesive under specific environmental conditions. This study found that the 793® adhesive offers the highest PPS for almost all $R_a$ studied. It is evident in Figure 11 that within the 0.17–0.21 μm range, the benchmark adhesive 496® and the second-best adhesive 401® exhibit similar behavior, though outside this specific range, 496® clearly surpasses 401®.

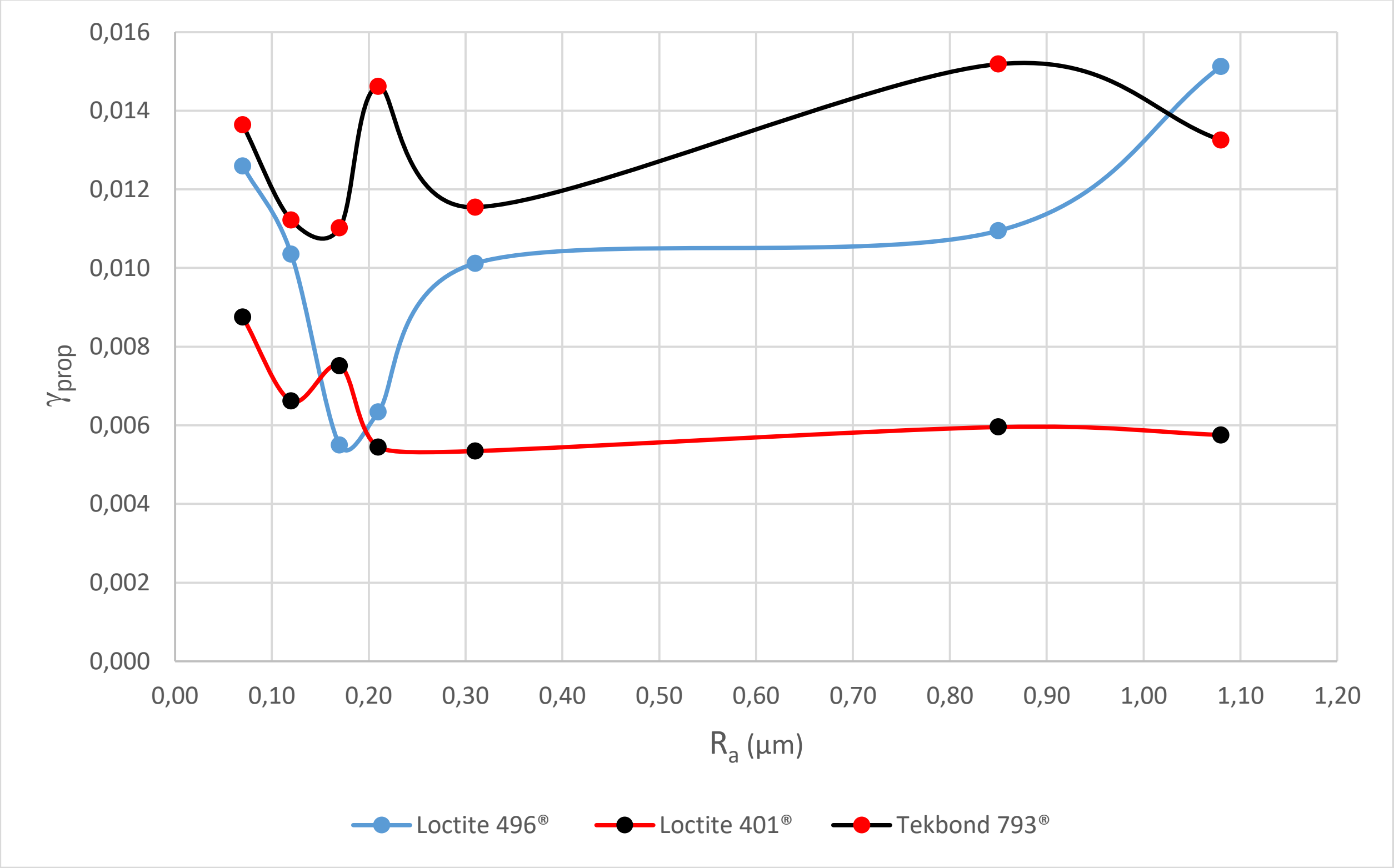

**Fig. 11.** Proportional Shear Strain x $R_a$

Figure 12 shows these findings in terms of sandpaper grit. It should also be highlighted that all three adhesives exceed a PSS of 0.005 or 5000 με across all $R_a$, which is well above the linear limit of most strain gages and even that of specialized gages.

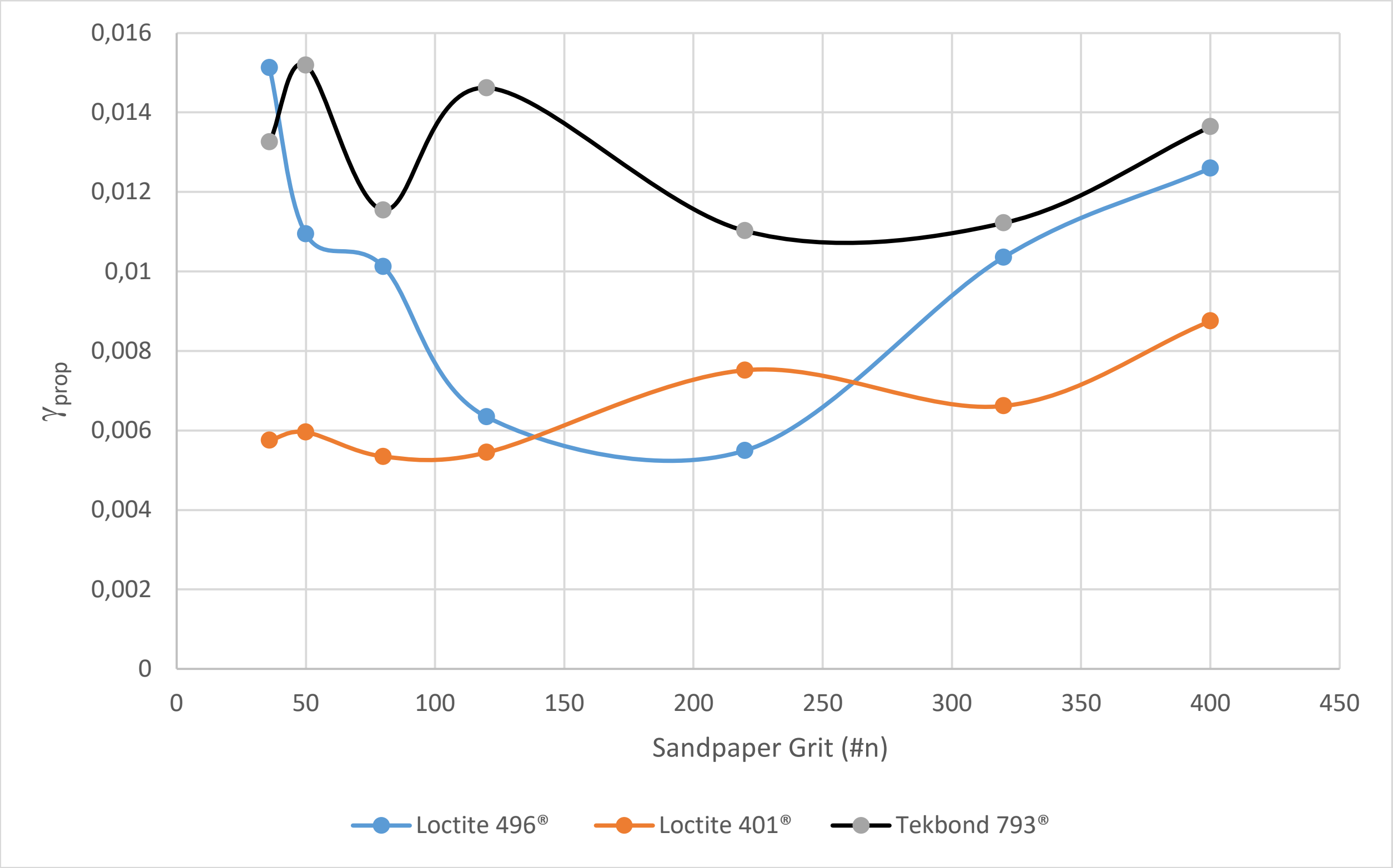

**Fig. 12.** Proportional Shear Strain x Sandpaper grit

### *3.5 Proportional Shear Stress versus $R_a$ and Grit*

The Proportional Shear Stress (PSSt) indicates the maximum point to which the adhesive can be applied without loss of linearity during tests. Once again, Tekbond 793® outperforms all other tested adhesives. Considering the typical sensor bonding surface roughness, 401® and 496® rank second and third respectively, as evidenced in Figures 13 and 14.

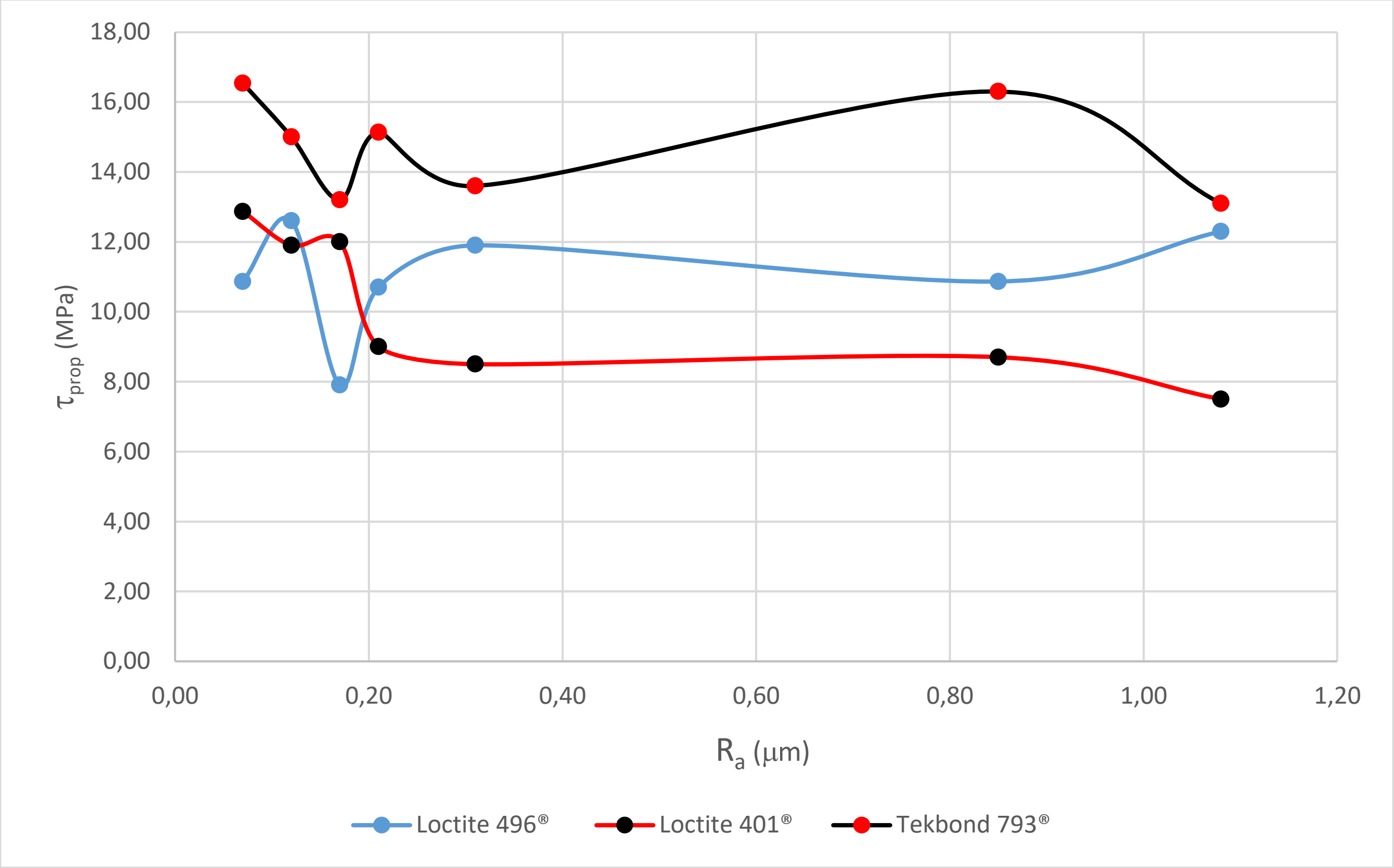

**Fig. 13.** Proportional Shear Stress x $R_a$

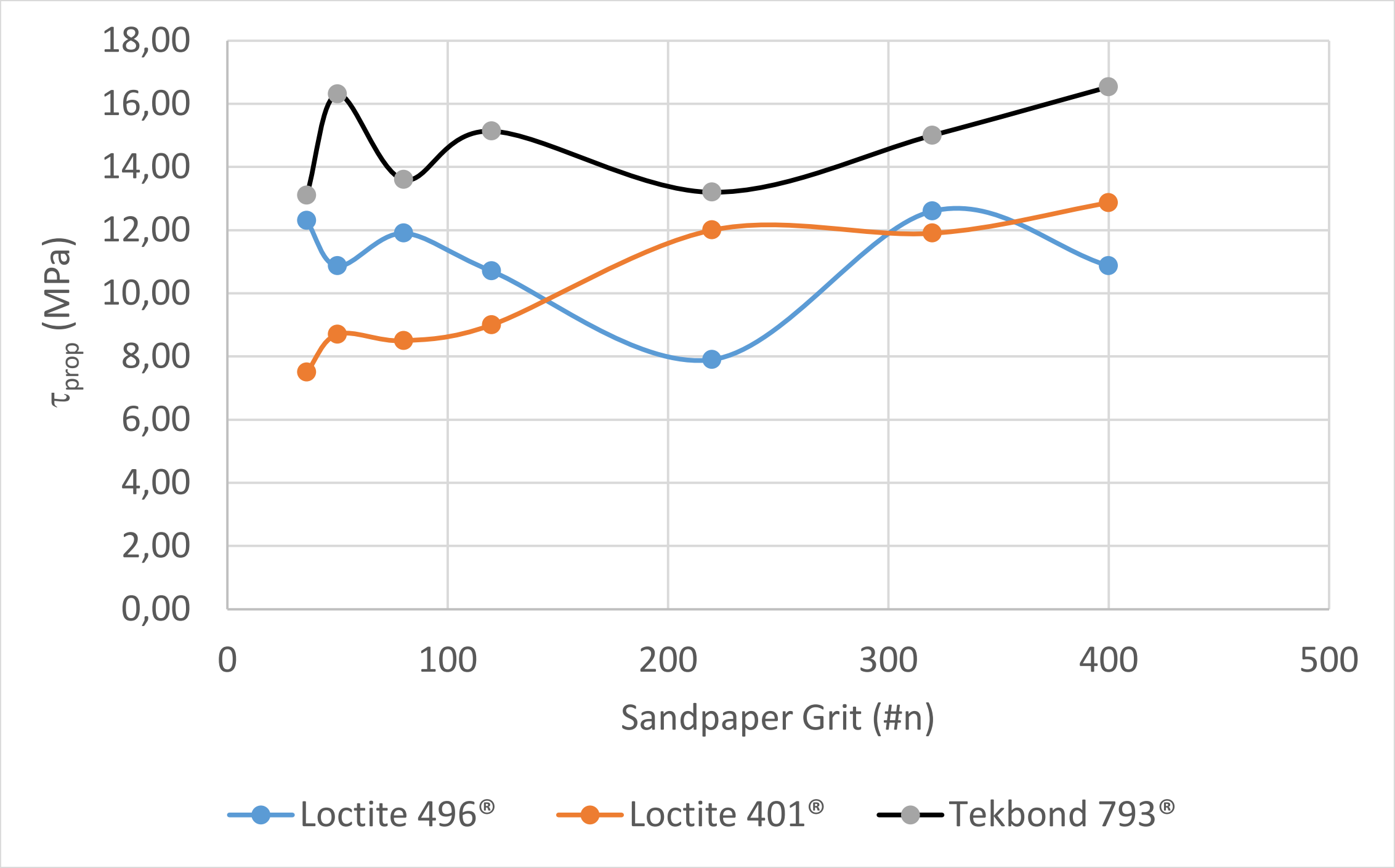

**Fig. 14.** Proportional Shear Stress x Sandpaper Grit

*3.6 Shear Strain at Maximum Shear Stress* (SSMSt) *versus $R_a$ and Grit*

The strain at the Maximum Shear Stress (MSSt), or SSMSt, is observed in two distinct regions as seen previously. Between a surface roughness ($R_a$) of 0.31 and 1.08 µm, the 496® adhesive exhibits the largest deformation which aligns with the fact that this adhesive has the lowest measured G* and Proportional Shear Strain (PPS). The 793® and 401® adhesives present very similar values, with 401® yielding the lowest SSMSt. However, the behavior is less clear at lower surface roughness (Ra) values. Figures 15 and 16 display the collected data.

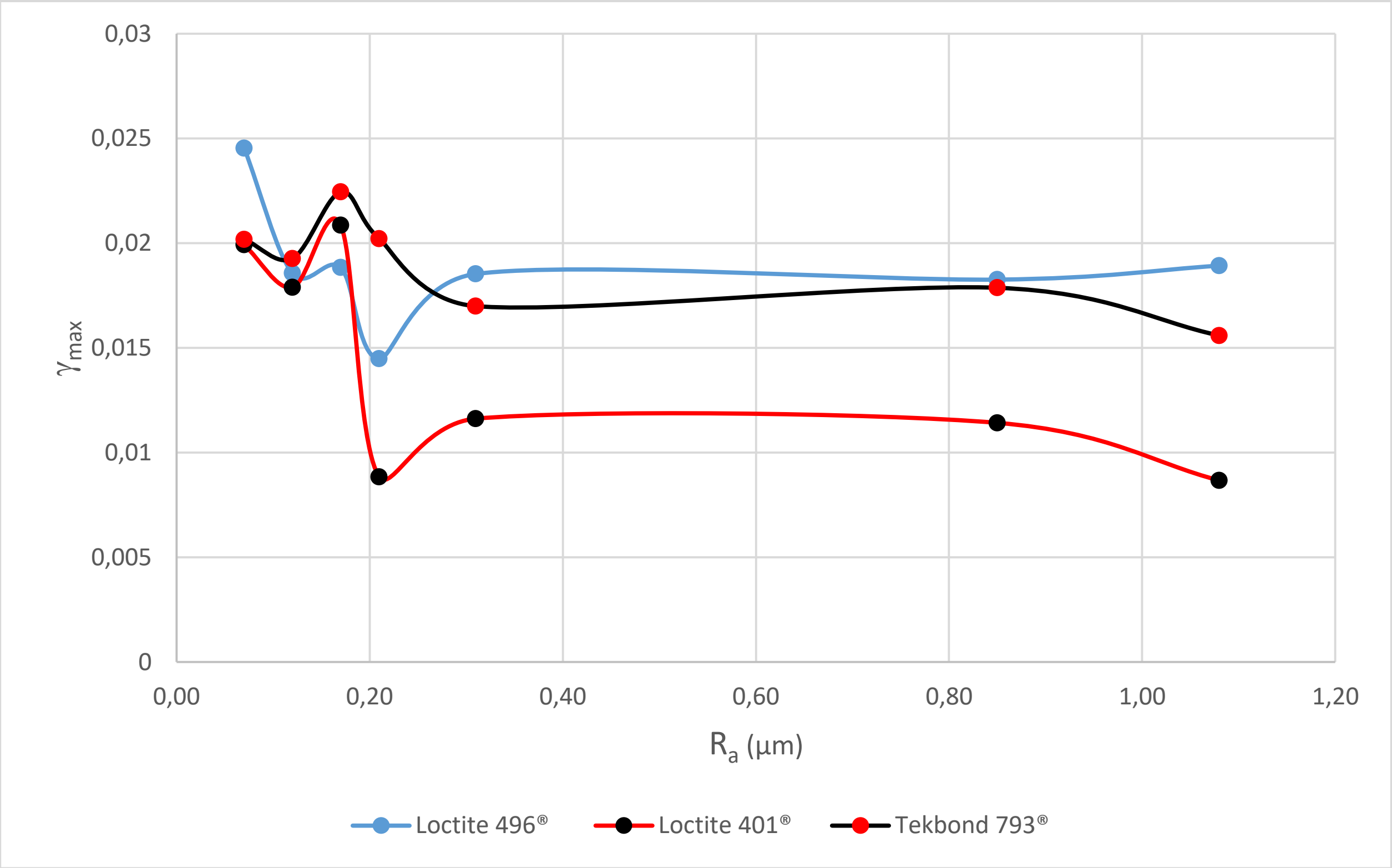

**Fig. 15.** Shear Strain at Maximum Shear Stress x $R_a$

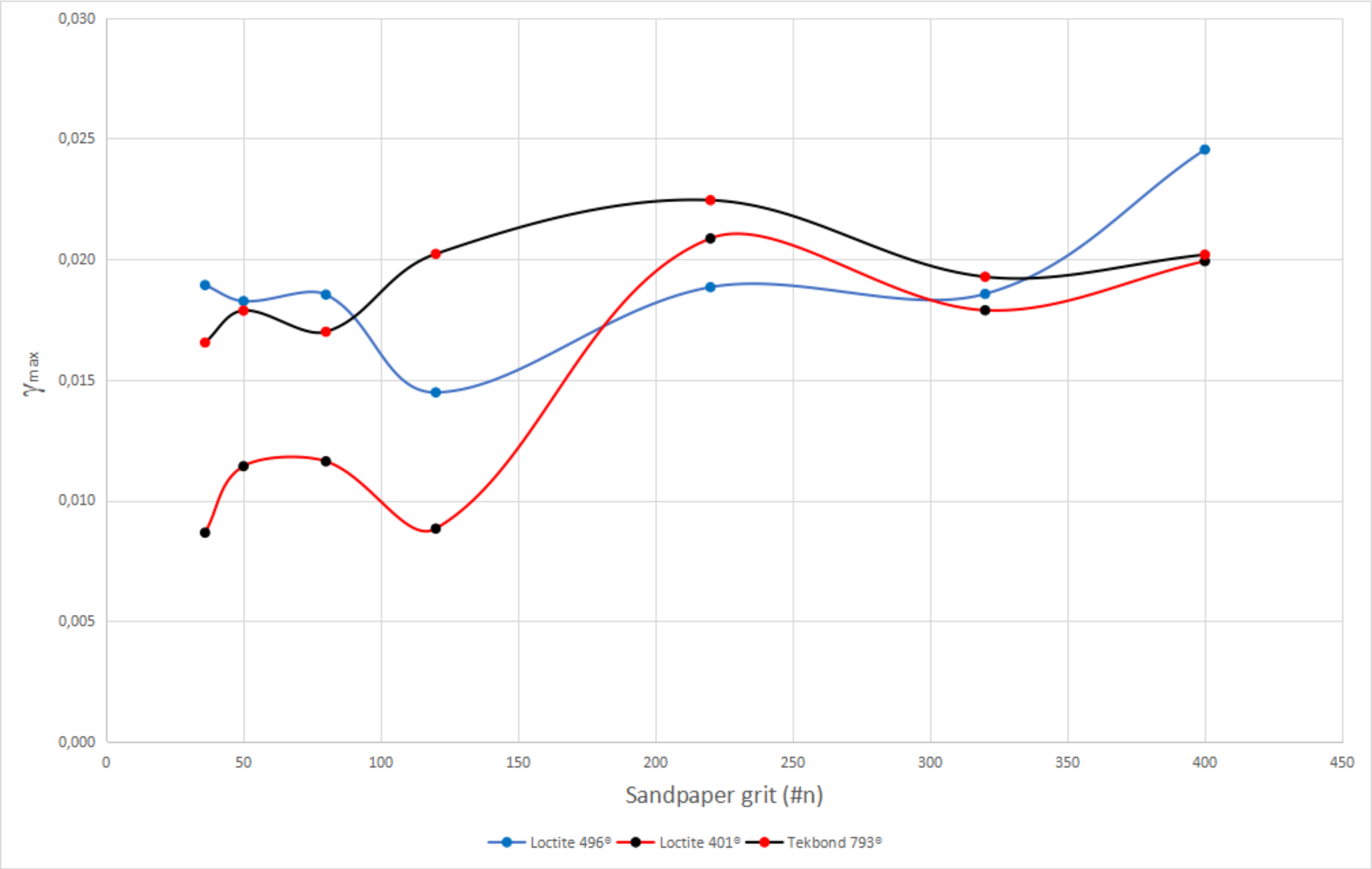

**Fig. 16.** Shear Strain at Maximum Shear Stress x Sandpaper Grit

### *3.7 Maximum Shear Stress versus $R_a$ and Grit*

In most cases, coarser grits tended to exhibit a consistent impact on the peak stress or Maximum Shear Stress (MSSt). For the finer sandpaper grits, the positions of both Loctite products shifted, as shown in Figure 17. The Tekbond glue exhibited superior values throughout all tests, with 496® taking second place (Figures 17 and 18). All values recorded were below those listed by references [4,5,6], yet higher than those reported in the literature [15-17].

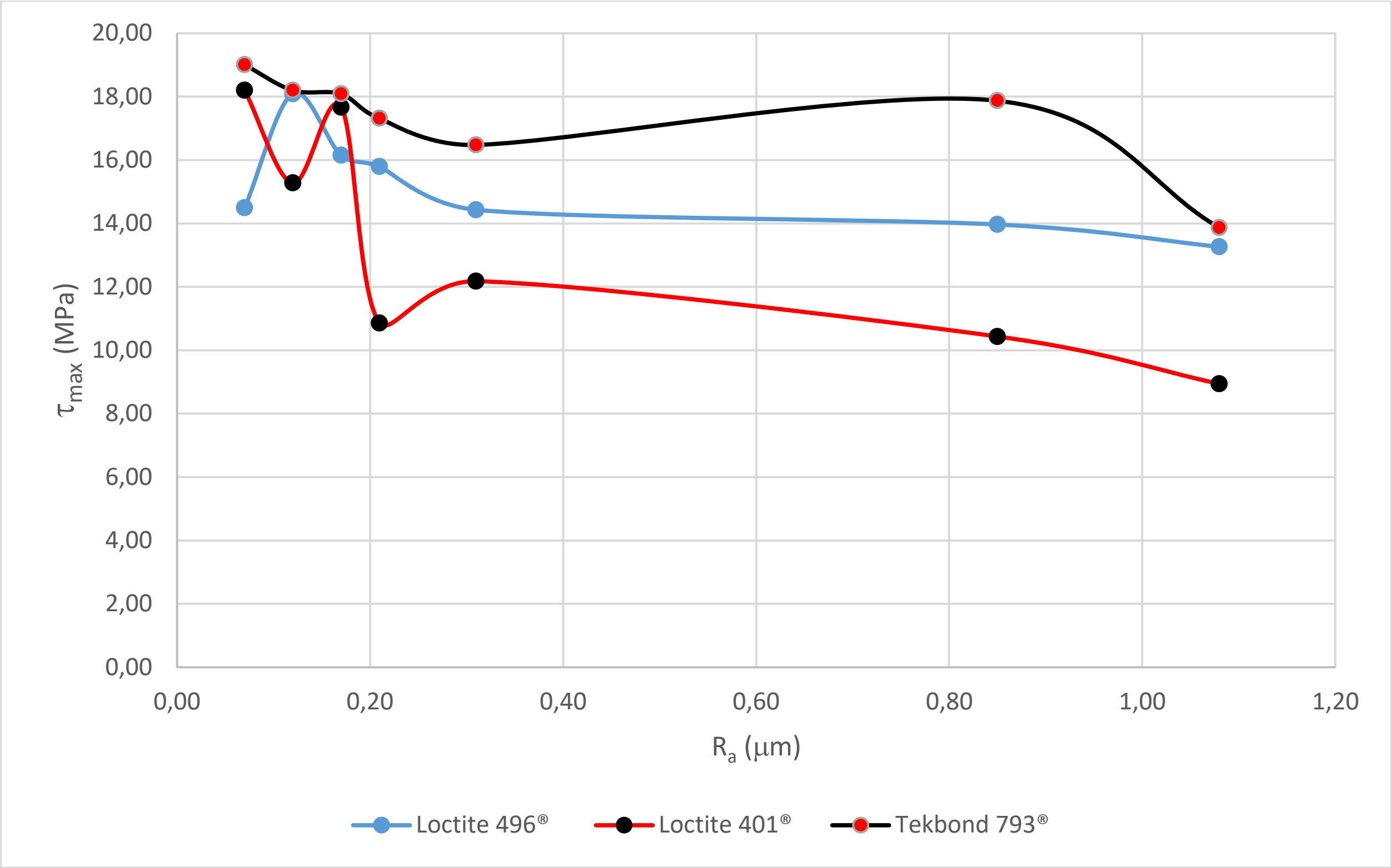

**Fig. 17.** Maximum Shear Stress x $R_a$

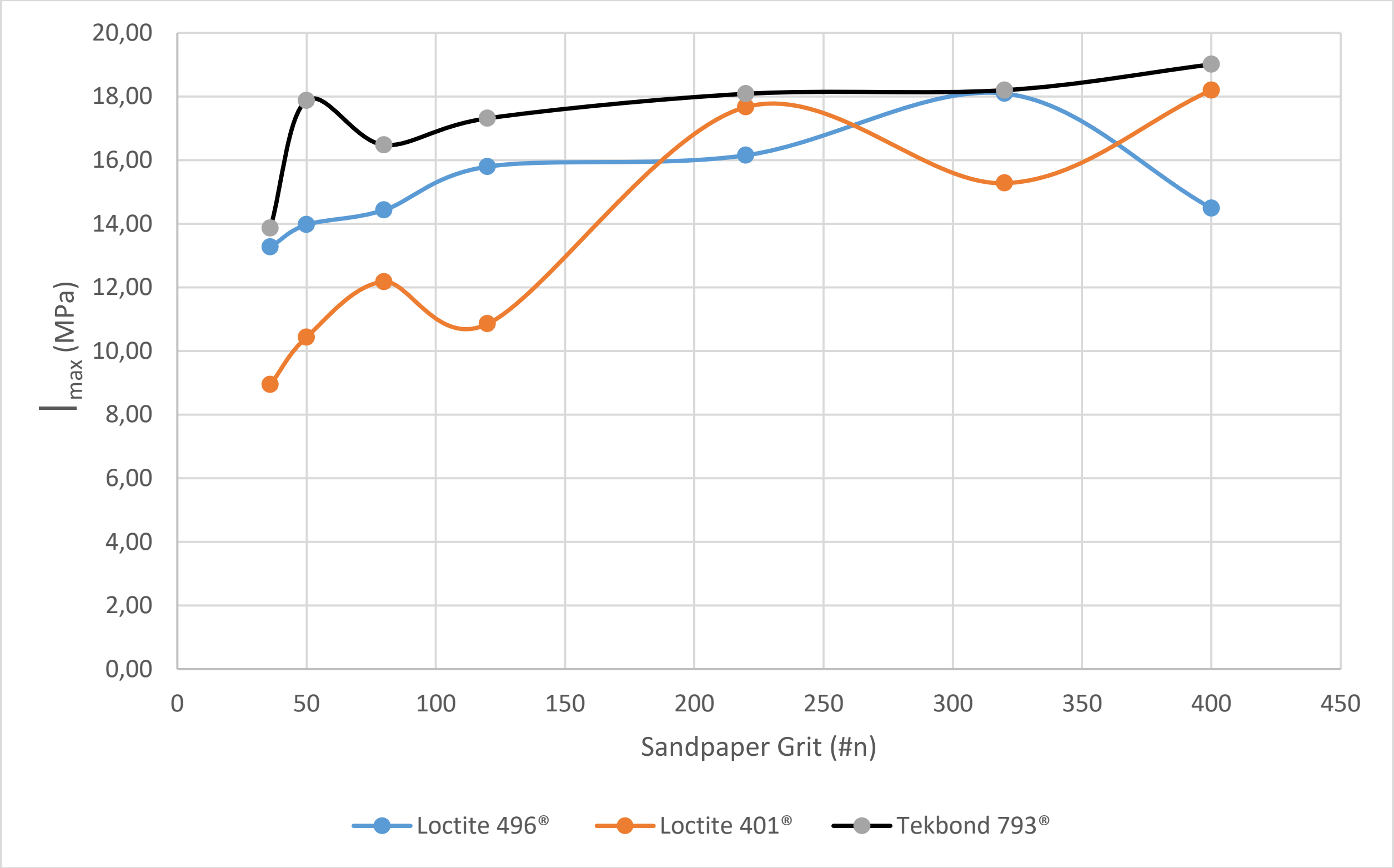

**Fig. 18.** Maximum Shear Stress x Sandpaper Grit (#n).

*3.8 Rupture Shear Strain versus $R_a$ and Grit*

The Rupture Shear Strain (RSS), while not typically a primary concern for strain gage users in comparison to linear properties and yield point determination, is nevertheless recorded for safety reasons. This ensures that portable sensors are not exposed to unnecessary risks.

In this study, the usual pattern of gradual and predictable strain variation was observed for higher $R_a$ values. On the contrary, a sinusoidal behavior prevailed for the lower $R_a$ values. Notwithstanding these trends, the benchmark 496® consistently demonstrated the greatest strain capability up to rupture, followed by 401®. For higher $R_a$ values, 793® exhibited the lowest rupture strain (see Figures 19 and 20).

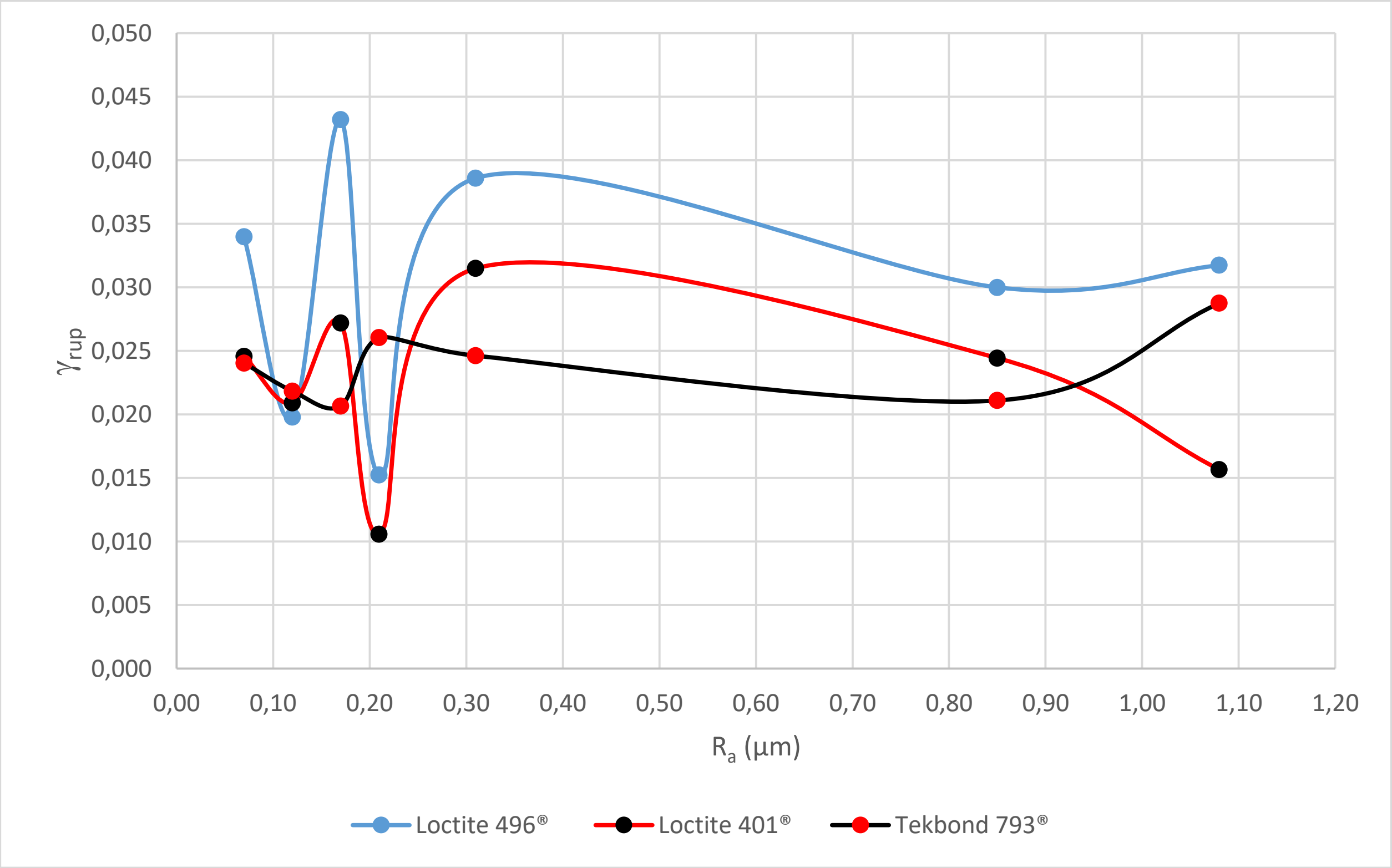

**Fig. 19.** Rupture Shear Strain x $R_a$

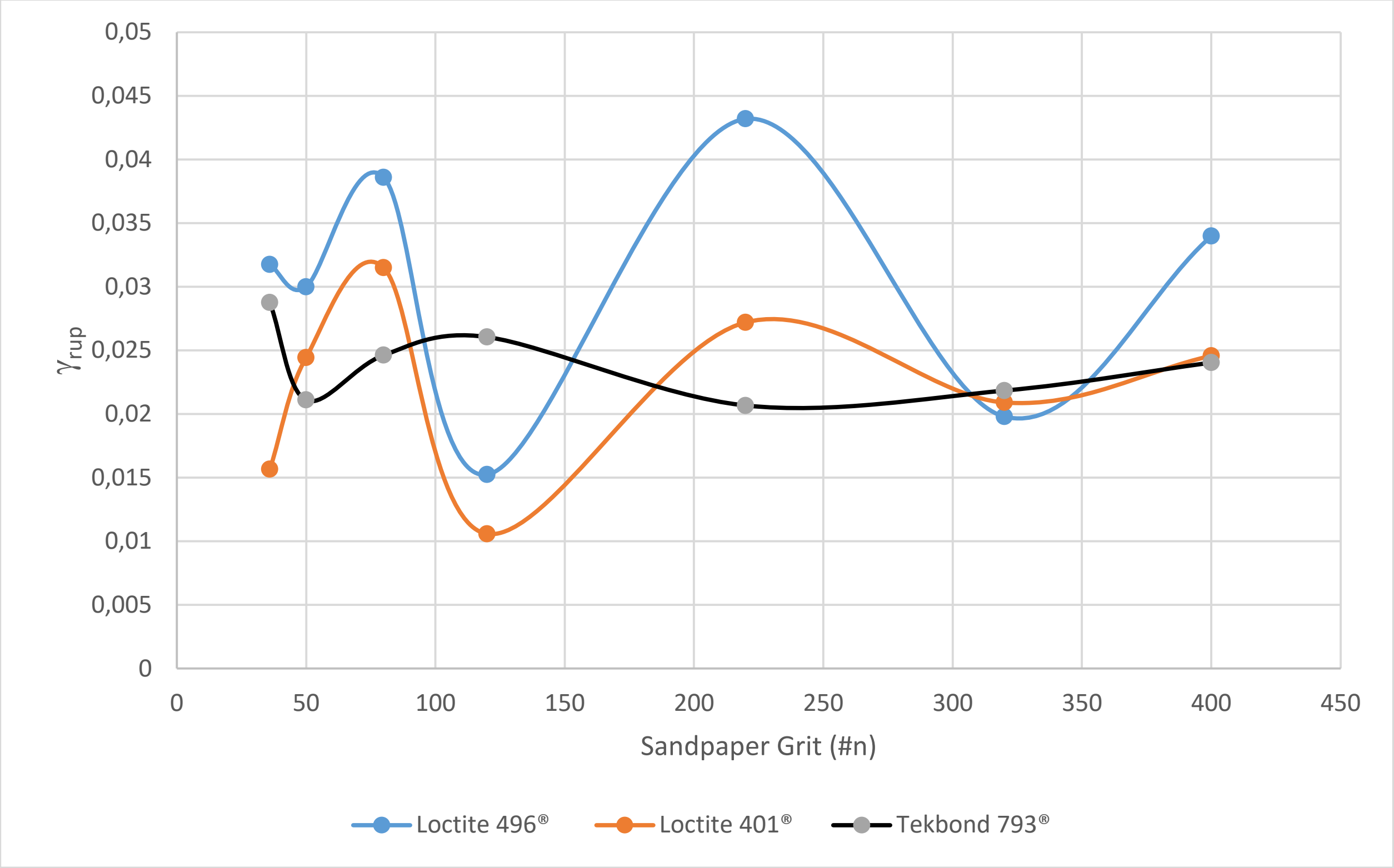

**Fig. 20.** Rupture Shear Strain x Sandpaper Grit

### *3.9 Rupture Shear Stress versus $R_a$ and Grit*

The Rupture Shear Stress (RSSt) data confirms that Tekbond 793® consistently exhibits the highest values for this property across nearly the entire range of $R_a$. This glue, with an average rupture shear stress of 15 MPa at $R_a$ values above 0.31 μm, should be considered in scenarios where there is a risk of setup rupture, sensor ejection, and damage, thereby ensuring hardware protection. Loctite 496® is the second-best performer with an average of 11.5 MPa, while Loctite 401® is at the lower end with 8 MPa. As the $R_a$ decreases, all three glues enhance their performance, as observed in Figures 21 and 22. Particularly, Tekbond 793® steadily increases its RSSt level up to just over 18 MPa at 0.07 μm or #400 (refer to Appendix C for exact values). While Loctite 496® demonstrates significant variability at lower $R_a$, Loctite 401® steadily improves and could potentially replace Loctite 496® if necessary, albeit considering only this aspect of adhesive selection.

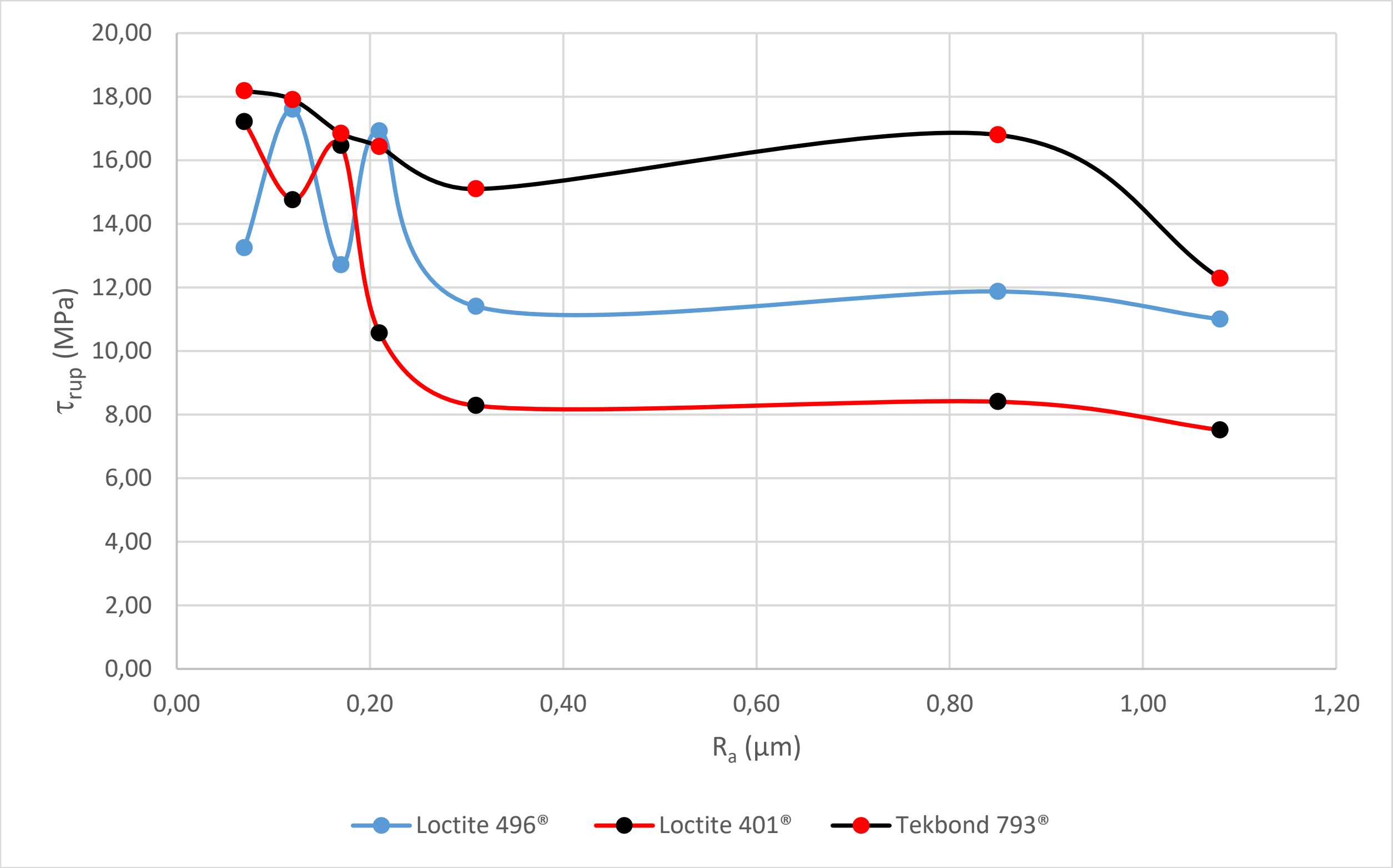

**Fig. 21.** Rupture Shear Stress x $R_a$

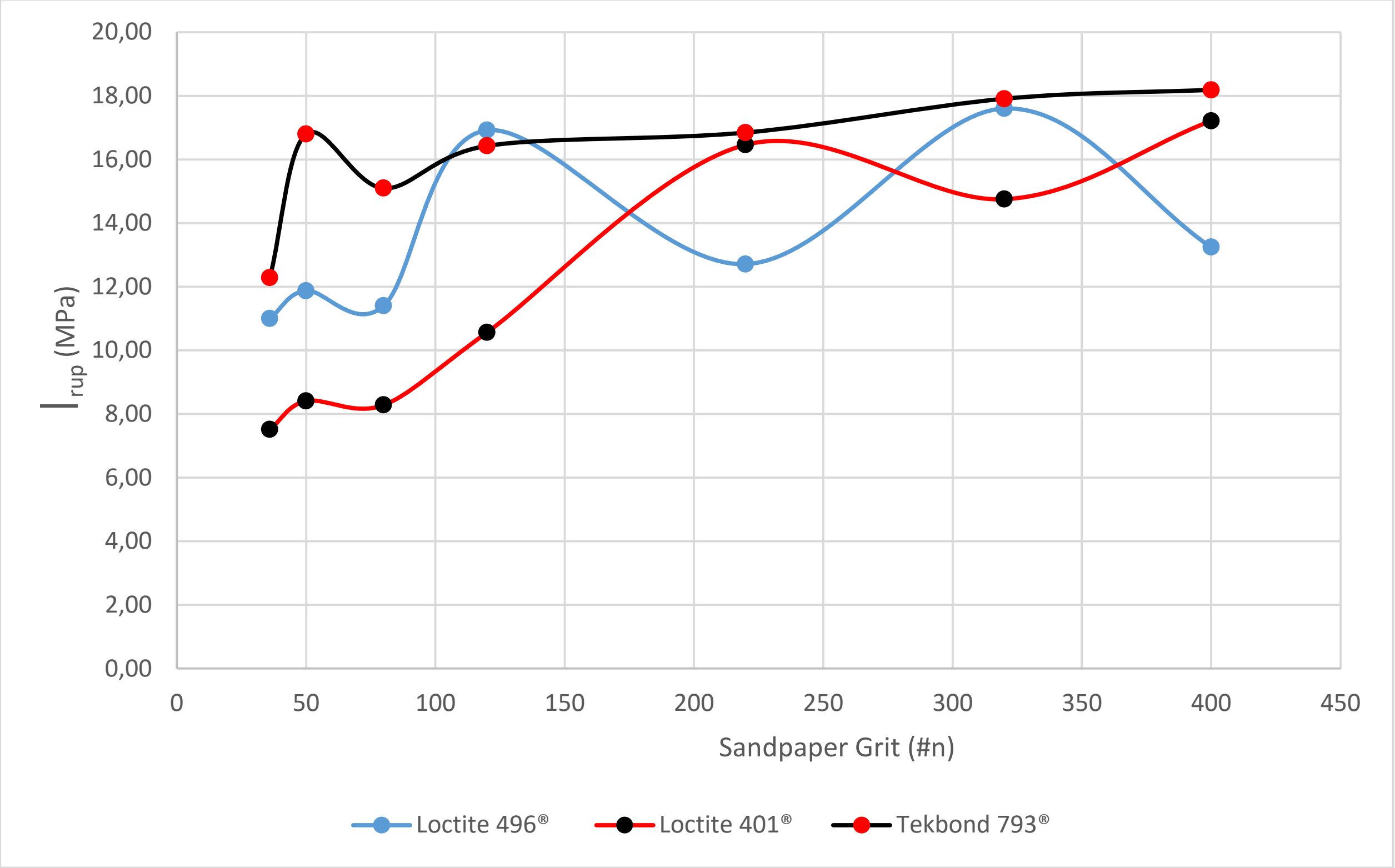

**Fig. 22.** Rupture Shear Stress x Sandpaper Grit

*3.10 Surface and Chemical Enhancement Effect Over Shear Stress and Strain*

To evaluate surface and chemical enhancement effects, two distinct investigations were conducted. The first examined the effects of the proposed surface treatment in isolation and in combination with a commercial accelerator on the benchmark Loctite 496® at a higher roughness of 0.85 μm. This roughness was achieved using #50 sandpaper and a random circular orbital sander. This grit number was selected as it falls within the linear variation zone previously discussed, and this roughness level is frequently employed for specific types of measurements (refer to Table 1). Following the abrasive process and prior to the application of any chemical enhancements, the surface was thoroughly cleaned as described in Section 2.1. Figure 23 presents the shear stress versus shear strain curves for these four surface conditions: simple abrasion (blue), abrasion plus Turbo Primer® (black), conditioning and neutralizing (red), and the combination of all treatments (yellow).

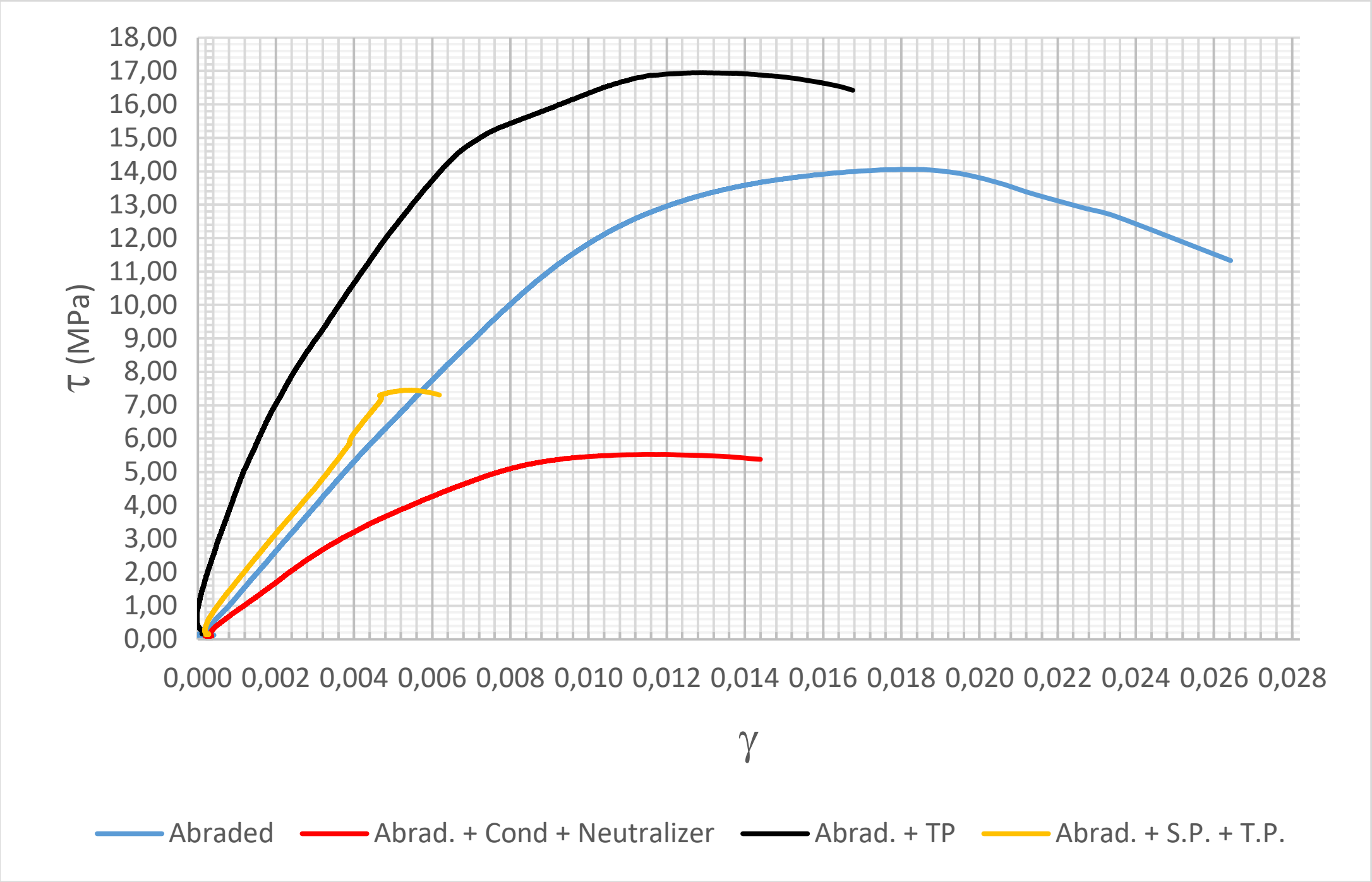

**Fig. 23.** Shear stress x strain curves for all tested surface conditions, Loctite 496. Surface was abraded with #50 sandpaper, producing an average of 0.85 μm.

The simple abraded surface exhibited the largest rupture or total shear strain limit and linear region among all cases. The use of the Turbo Primer® accelerator improved the Maximum Shear Stress but resulted in a decrease in rupture strain. The accelerator enhanced stiffness (represented by G*) compared to all other options. Conversely, the use of chemicals to remove oxidation and subsequent neutralization resulted in the lowest stress-strain curve, reducing the Proportional Shear limit and the Maximum Shear Stress value. Upon adding the accelerator, stiffness and proportional shear were improved, but at the expense of total shear strain. Figure 23 demonstrates that the accelerator induced the formation of a linear region, similar to the simple abraded case. Lastly, for this case, Table 6 presents the maximum stresses and strains for comparison purposes.

**Table 6**
Some mechanical properties for the four studied surface treatment using Loctite 496®, Grit sandpaper = #50 and resulting $R_a$ = 0.85 μm.

| Surface Treatment | Maximum Shear Stress (MPa) | Prop. Shear Stress (MPa) | Prop Shear Strain | G* (MPa) |
|---|---|---|---|---|
| Abraded | 14,05 | 9,80 | 0,021807 | 1089 |
| Abr + TP® | 16,95 | 14,00 | 0,006003 | 3258 |
| Abr + Cond + Neutr | 5,53 | 3,00 | 0,0034651 | 757 |
| Abr + Cond + Neutr + TP® | 7,45 | 7,20 | 0,0044966 | 1397 |

In contrast, the second study was conducted using a different grit, #320, which corresponds to a $R_a$ of 0.12 μm and falls within the zone of high variation. This sandpaper grit is also highlighted in literature as representing the high end [1] for optimal bonding results. Figure 24 displays the outcomes for the three glues under these conditions.

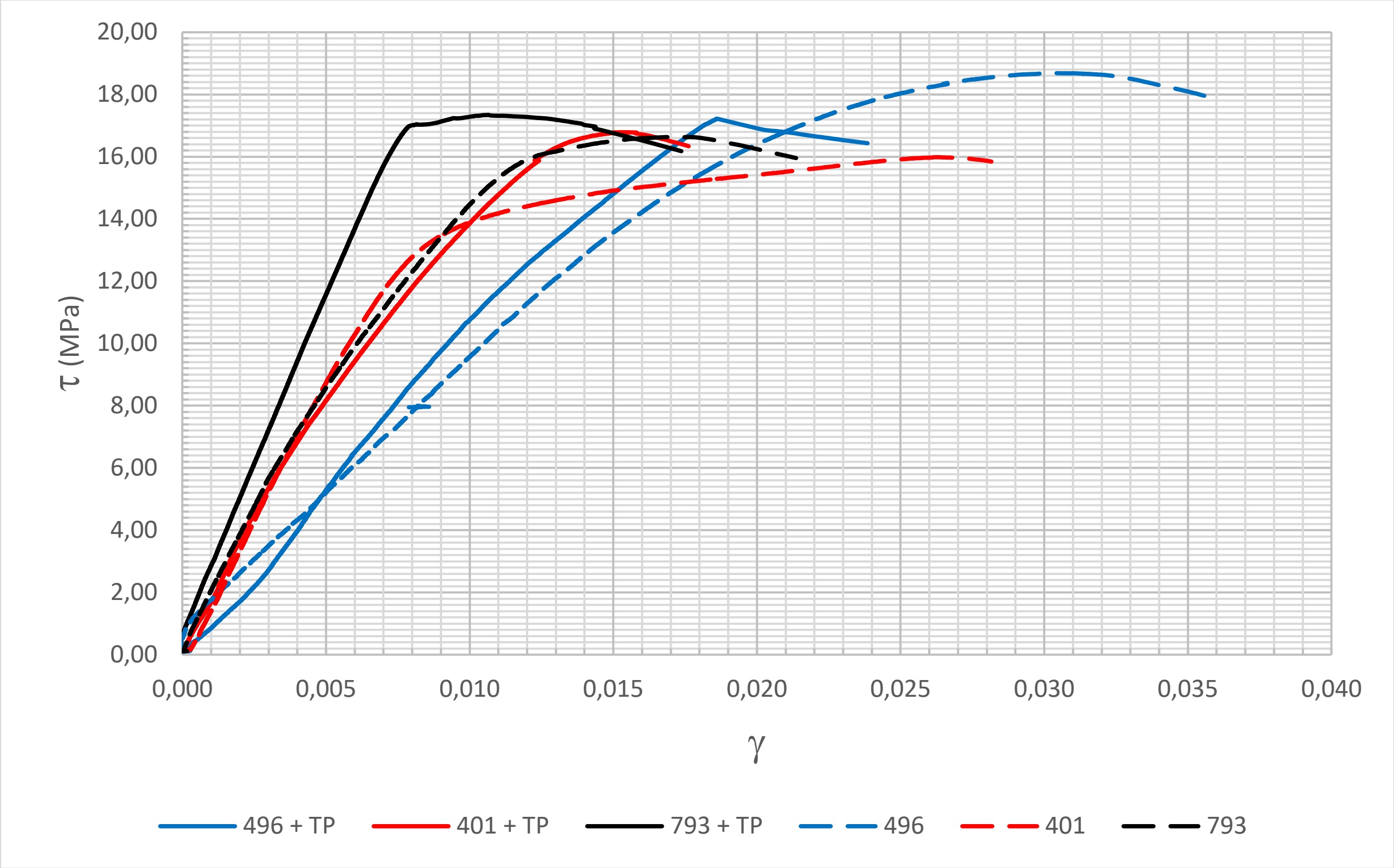

**Fig. 24.** Accelerator (Turbo Primer®) effect over the three studied glues.

When used alone, the benchmark adhesive 496® produced a typical stress-strain curve (dashed line). The accelerator promoted faster initial adhesion and stiffer joint (higher G*), but it also lowered maximum strength and total strain. With regards to 401®, the use of Turbo Primer® did not reduce the initial adhesion time as much as in the benchmark case. Although it slightly reduced the stiffness, it improved the maximum shear stress. The total strain was reduced as well, but this may not constitute a crucial feature for the sensor user. A similar pattern was observed for Tekbond 793®. Table 7 provides some of the data collected from these tests.

**Table 7**
Studied cases comparing the effect of an accelerator over some mechanical properties. #320 and 0.12 µm.

| | 496® | | 401® | | 793® | |
|---|---|---|---|---|---|---|
| | A | A+TP | A | A+TP | A | A+TP |
| Max. Shear Stress (MPa) | 18.68 | 17.22 | 15.98 | 16.78 | 16.63 | 19.51 |
| Prop. Shear Stress (MPa) | 12.80 | 16.80 | 11.80 | 15.80 | 13.80 | 19.00 |
| Prop. Shear Strain | 0.013932 | 0.017817 | 0.007080 | 0.012292 | 0.009352 | 0.015406 |
| G* (MPa) | 865 | 1269 | 1824 | 1704 | 1771 | 2193 |

## 4. Conclusions

It is not feasible to directly simulate the behavior of strain gage adhesive using two steel plates, given the nature of strain gages as thin metallic foils constructed on a polymeric backing. Consequently, the most effective adhesive simulation is indirect, comparing to a widely accepted glue whose performance is considered satisfactory by stress analysts. This approach allows for a relative performance evaluation when other adhesives are subjected to the same conditions. The behavior of the replacement benchmark, Loctite 496®, was observed and measured, and also potential replacements, Loctite 401® and Tekbond 793®, were tested and compared. For strain gage bonding, the most pertinent mechanical properties are Proportional Shear Strain (PSS), its consequent Proportional Shear Stress (PSSt), and the Apparent Shear Modulus (G*). Long-term or cyclic properties were not evaluated in this study.

Loctite 401® yielded the highest G*, suggesting that it should be used during strain gage installation if this property is paramount. On the other hand, Tekbond 793® exhibited the highest Proportional Shear Strain (PSS) among the three adhesives, revealing very linear behavior, even without an accelerator. Loctite 496® achieved the highest Rupture Shear Strain (RSS) and, in tandem with Tekbond 793®, the highest Shear Strain at MSSt.

Generally, the use of an accelerator is recommended for all tested adhesives. However, the application of a surface conditioner and neutralizer significantly reduced adhesive performance.

The study found that hand abrading produced nearly double the roughness created by an orbital sander, a factor that must be considered by stress analysts when preparing surfaces for study.

Additionally, two distinct regions are consistently observed in the resulting graphs relating $R_a$ with a given mechanical property. The first is a highly dispersive region, wherein $R_a$ increases from 0.07 μm to 0.31 μm, and a given property changes significantly. Thereafter, when $R_a$ exceeds 0.31 μm, a plateau is seen up to 1.05 μm. While several works focus on higher $R_a$ levels with an emphasis on optimizing bonding properties over sensor performance, this study also opted to examine $R_a$ values generated by common strain gage surface preparation. The significant variation observed in mechanical properties at lower $R_a$ values (below 0.31 μm) suggests that factors beyond adhesive layer thickness influence the bond, not only the adhesive layer thickness as it has been pointed out. Such variations are less significant for $R_a$ values above 0.31 μm.

Finally, all three glues tend to fail in the midsection of their thickness, with a regular variation across the failed area.

# APPENDIX A

**Table A.1: Adhesive layer thickness statistical results**

| Glue | Grit # | Avg 1 | Std 1 | Avg2 | Std 2 | spc | thickness | Average (µm) |
|---|---|---|---|---|---|---|---|---|
| 496 | 320 | 13 | 5,6 | 14 | 5,9 | 1, 2 e 3 | 27 | **29** |
| 496 + TP | 320 | 19 | 7,2 | 15 | 5,1 | 1, 2 e 3 | 34 | **34** |
| 401 | 320 | 23 | 10,8 | 16 | 9,1 | 4, 5, e 6 | 39 | |
| 401 | 220 | 15 | 8,3 | 19 | 11,4 | 1, 2 e 3 | 34 | **39** |
| 401 | 400 | 22 | 11,2 | 18 | 12,8 | 4, 5 e 6 | 40 | |
| 401 + TP | 320 | 20 | 13,6 | 21 | 11,9 | 1, ,2 e 3 | 41 | **41** |
| TK 793 | 320 | 15 | 5,8 | 15 | 7,2 | 4, 5 e 6 | 31 | |
| TK 793 | 400 | 14 | 3,6 | 13 | 3,8 | 4, 5 e 6 | 28 | **29** |
| TK 793 + TP | 400 | 19 | 7,0 | 19 | 7,5 | 1, 2 e 3 | 38 | **38** |

## APPENDIX B

**Table B.1: Sandpaper grit and roughness measured values**

| Grit # = | 36 | 50 | 80 | 120 | 220 | 320 | 400 |
|---|---|---|---|---|---|---|---|
| **1** | 1,18 | 0,85 | 0,43 | 0,17 | 0,16 | 0,14 | 0,07 |
| **2** | 1,18 | 0,78 | 0,41 | 0,18 | 0,20 | 0,12 | 0,07 |
| **3** | 1,04 | 0,88 | 0,31 | 0,19 | 0,17 | 0,13 | 0,07 |
| **4** | 1,15 | 0,81 | 0,31 | 0,19 | 0,16 | 0,12 | 0,06 |
| **5** | 1,13 | 0,84 | 0,34 | 0,21 | 0,21 | 0,15 | 0,07 |
| **6** | 0,99 | 0,76 | 0,35 | 0,22 | 0,17 | 0,13 | 0,07 |
| **7** | 0,90 | 0,95 | 0,35 | 0,19 | 0,19 | 0,10 | 0,07 |
| **8** | 1,17 | 0,94 | 0,35 | 0,19 | 0,16 | 0,10 | 0,08 |
| **9** | 1,00 | 0,90 | 0,36 | 0,22 | 0,15 | 0,11 | 0,06 |
| **10** | 1,12 | 0,98 | 0,31 | 0,19 | 0,16 | 0,14 | 0,07 |
| **11** | 1,07 | 0,79 | 0,30 | 0,21 | 0,17 | 0,13 | 0,08 |
| **12** | 1,00 | 0,90 | 0,28 | 0,20 | 0,16 | 0,15 | 0,07 |
| **13** | 0,91 | 0,95 | 0,30 | 0,19 | 0,14 | 0,12 | 0,06 |
| **14** | 1,24 | 0,83 | 0,28 | 0,19 | 0,17 | 0,14 | 0,07 |
| **15** | 1,08 | 0,95 | 0,29 | 0,22 | 0,18 | 0,12 | 0,07 |
| **16** | 1,25 | 0,70 | 0,31 | 0,21 | 0,16 | 0,12 | 0,07 |
| **17** | 1,17 | 0,66 | 0,34 | 0,23 | 0,16 | 0,11 | 0,06 |
| **18** | 1,20 | 0,61 | 0,32 | 0,19 | 0,17 | 0,11 | 0,08 |
| **19** | 1,12 | 0,71 | 0,32 | 0,22 | 0,17 | 0,12 | 0,08 |
| **20** | 1,07 | 0,82 | 0,35 | 0,24 | 0,16 | 0,11 | 0,06 |
| **21** | 1,16 | 0,90 | 0,25 | 0,22 | 0,17 | 0,12 | 0,06 |
| **22** | 1,09 | 0,94 | 0,24 | 0,20 | 0,17 | 0,11 | 0,08 |
| **23** | 0,97 | 0,86 | 0,22 | 0,26 | 0,18 | 0,10 | 0,06 |
| **24** | 0,89 | 0,92 | 0,22 | 0,20 | 0,18 | 0,09 | 0,07 |
| **25** | 1,06 | 0,87 | 0,30 | 0,23 | 0,17 | 0,11 | 0,07 |
| **26** | 1,02 | 0,92 | 0,33 | 0,23 | 0,14 | 0,15 | 0,06 |
| **27** | 1,02 | 0,90 | 0,33 | 0,23 | 0,14 | 0,12 | 0,07 |
| **28** | 1,03 | 0,98 | 0,28 | | 0,16 | 0,12 | |
| **29** | 1,06 | 0,81 | 0,26 | | 0,19 | 0,11 | |
| **30** | 1,08 | 0,85 | 0,27 | | 0,14 | 0,13 | |
| **Avg =** | **1,08** | **0,85** | **0,31** | **0,21** | **0,17** | **0,12** | **0,07** |
| **Std Dev =** | **0,10** | **0,09** | **0,05** | **0,02** | **0,02** | **0,02** | **0,01** |

| 95% Conf Int = | **0,19** | **0,18** | **0,10** | **0,04** | **0,03** | **0,03** | **0,01** |
|---|---|---|---|---|---|---|---|

## APPENDIX C

**Table C.1:**
Overall results for this study

| | Loctite 496® | | | | | | | |
|---|---|---|---|---|---|---|---|---|
| Ra (µm) | Sdpp Grit (#n) | G (MPa) | τmax (MPa) | γ@max | τprop (MPa) | γprop | τrup (MPa) | γrupt |
| 1,08 | 36 | 871 | 13,26 | 0,0189209 | 12,30 | 0,015124 | 11,00 | 0,031734 |
| 0,85 | 50 | 918 | 13,97 | 0,0182547 | 10,87 | 0,010946 | 11,87 | 0,029996 |
| 0,31 | 80 | 1221 | 14,43 | 0,0185259 | 11,90 | 0,010121 | 11,40 | 0,038591 |
| 0,21 | 120 | 1402 | 15,79 | 0,0144737 | 10,70 | 0,00634 | 16,92 | 0,015239 |
| 0,17 | 220 | 1430 | 16,15 | 0,0188334 | 7,90 | 0,005497 | 12,71 | 0,043184 |
| 0,12 | 320 | 1322 | 18,09 | 0,0185557 | 12,60 | 0,010356 | 17,60 | 0,019798 |
| 0,07 | 400 | 921 | 14,49 | 0,0245364 | 10,87 | 0,012594 | 13,24 | 0,033964 |

| | Loctite 401® | | | | | | | |
|---|---|---|---|---|---|---|---|---|
| Ra (µm) | Sdpp Grit (#n) | G (MPa) | τmax (MPa) | γ@max | τprop (MPa) | γprop | τrup (MPa) | γrupt |
| 1,08 | 36 | 1490 | 8,94 | 0,0086636 | 7,50 | 0,005756 | 7,52 | 0,015657 |
| 0,85 | 50 | 1575 | 10,43 | 0,0114165 | 8,70 | 0,005956 | 8,41 | 0,024434 |
| 0,31 | 80 | 1775 | 12,18 | 0,0116171 | 8,50 | 0,005346 | 8,28 | 0,031492 |
| 0,21 | 120 | 2101 | 10,85 | 0,0088292 | 9,00 | 0,005446 | 10,57 | 0,010583 |
| 0,17 | 220 | 2010 | 17,67 | 0,0208535 | 12,00 | 0,007516 | 16,46 | 0,027203 |
| 0,12 | 320 | 1979 | 15,27 | 0,0178783 | 11,90 | 0,006616 | 14,75 | 0,020903 |
| 0,07 | 400 | 1444 | 18,20 | 0,0199124 | 12,87 | 0,00875 | 17,21 | 0,024572 |

| | Tekbond 793® | | | | | | | |
|---|---|---|---|---|---|---|---|---|
| Ra (µm) | Sdpp Grit (#n) | G (MPa) | τmax (MPa) | γ@max | τprop (MPa) | γprop | τrup (MPa) | γrupt |
| 1,08 | 36 | 697 | 15,05 | 0,016540 | 14,40 | 0,014762 | 12,28 | 0,028747 |
| 0,85 | 50 | 1080 | 17,87 | 0,017865 | 16,30 | 0,015185 | 16,80 | 0,021096 |
| 0,31 | 80 | 1226 | 16,48 | 0,016987 | 13,60 | 0,011547 | 15,09 | 0,024617 |
| 0,21 | 120 | 1116 | 17,31 | 0,020212 | 15,13 | 0,014620 | 16,43 | 0,026046 |
| 0,17 | 220 | 1263 | 18,08 | 0,022439 | 13,20 | 0,011020 | 16,84 | 0,020668 |
| 0,12 | 320 | 1674 | 18,20 | 0,019264 | 15,00 | 0,011219 | 17,91 | 0,021839 |
| 0,07 | 400 | 1664 | 19,01 | 0,020178 | 16,53 | 0,013639 | 18,19 | 0,024037 |